%% file: main.tex
\documentclass[prd,twocolumn,aps,letterpaper,amsmath,amssymb,preprintnumbers,showpacs,superscriptaddress,floatfix,longbibliography,nofootinbib]{revtex4-2}
\usepackage{array}
\usepackage{graphicx}
\usepackage[export]{adjustbox}
\usepackage[caption=false]{subfig}
\usepackage{tikz}
\usetikzlibrary{tikzmark}
\usetikzlibrary{calc}
\allowdisplaybreaks 
\usepackage[hypertexnames=true]{hyperref}
\usepackage[capitalize]{cleveref}
\Crefname{appsec}{appendix}{appendices}
\hypersetup{
    colorlinks=true,       
    linkcolor=blue,          
    citecolor=blue,        
    filecolor=blue,      
    urlcolor=blue           
}

\usepackage{slashed}
\usepackage{mathtools}
\usepackage{listings}
\usepackage{pgf}
\usepackage{fancyvrb}
\usepackage{longtable}
\usepackage{bm}
\usepackage{tikz-cd}
\usepackage{amsthm}
\usepackage{dsfont}
\usepackage{mathrsfs}
\usepackage{algorithm}
\usepackage{algpseudocode}
\usepackage{centernot}
\usepackage{booktabs}
\usepackage{colortbl}

\makeatletter
\renewcommand\p@subfigure[1]{\thefigure(#1)}
\makeatother

\DeclareMathOperator*{\argmin}{arg\,min}
\DeclareMathOperator{\imag}{Im}
\DeclareMathOperator{\real}{Re}
\DeclareMathOperator{\erf}{erf}
\newcommand{\R}[0]{\ensuremath{\mathbb{R}}}
\newcommand{\C}[0]{\ensuremath{\mathbb{C}}}
\newcommand{\N}[0]{\ensuremath{\mathbb{N}}}
\newcommand{\ktt}[0]{\ensuremath{\mathtt{k}}}
\newcommand{\gtt}[0]{\ensuremath{\mathtt{g}}}
\newcommand{\ctt}[0]{\ensuremath{\mathtt{c}}}

\newcommand{\regulator}{\ensuremath{\xi}}
\newcommand{\norm}[1]{\lVert #1 \rVert}
\newcommand{\abs}[1]{\ensuremath{| #1|}}
\newcommand{\order}[1]{\ensuremath{\mathcal{O}\left( #1 \right)}}

\definecolor{darkgreen}{rgb}{0.0, 0.545098, 0.0}

\usepackage[normalem]{ulem}

\begin{document}
\title{
Kernel transformations and bounds for smeared spectral functions
}
\begin{abstract}
This work develops a framework for transforming between smeared spectral functions computed using different smearing kernels.
The kernel-transformation problem naturally arises when information is available for one family of energy-smeared observables, while phenomenology or comparison with other calculations require a different smearing.
For exact transformations, analytic conditions are established for the maps to exist and converge without arbitrary regularization.
Explicit expressions are provided for several kernel classes of interest, including Cauchy-to-Gaussian transformations and Gaussian-to-Cauchy width mixtures.
When exact transformations are unavailable, the inverse problem is tackled through regulated maps paired with bounds on the associated systematic error, directly computable from the given input data.
Errors on the input smeared spectral functions, either statistical or in the form of pointwise rigorous bounds, are then propagated to the target observables.
Enforcing spectral positivity can be used to tighten the bounds.
\end{abstract}

\author{William~I.~Jay}
\email{william.jay@colostate.edu}
\affiliation{Department of Physics, Colorado State University, Fort Collins, CO 80523, USA}

\author{Matteo Saccardi}
\email{matteo.saccardi@colostate.edu}
\affiliation{Department of Physics, Colorado State University, Fort Collins, Colorado 80523, USA}

\date{\today}

\maketitle

\section{Introduction}
\input{sec_introduction.tex}

\section{Kernel transformations}\label{sec:formalism}
\input{sec_kernel_transformations.tex}

\section{Bounds and Asymptotics}\label{sec:bounds}
\input{sec_bounds.tex}

\section{Numerical Realization}\label{sec:numerics}
\input{sec_numerics.tex}

\section{Conclusions}
\input{sec_conclusions.tex}

\section*{A\lowercase{cknowledgements}}

We thank Ryan Abbott, Norman Christ, Sarah Fields, and Patrick Oare for useful discussions.
M.S. gratefully acknowledges useful conversations with Mattia Bruno on the subject of inverse problems in physics.

The numerical calculations in this work made use of \textsc{NumPy}~\cite{vanderWalt:2011bqk,Harris:2020xlr} and \textsc{mpmath}~\cite{mpmath}.
Figures were generated using \textsc{matplotlib}~\cite{Hunter:2007} and \textsc{seaborn}~\cite{Waskom2021}.
Convex optimization was performed using \textsc{cvxpy}~\cite{diamond2016cvxpy,agrawal2018rewriting} and \textsc{clarabel}~\cite{clarabel}.

\section*{Data availability}
Code for the numeric examples in Sec.~\ref{sec:numerics} is available at Ref.~\cite{code_github}.

\appendix
\renewcommand{\appendixname}{APPENDIX} 
\input{app_normalization.tex}
\input{app_rk.tex}
\input{app_sip_dual.tex}
\bibliography{main.bib}

\end{document}

%% file: sec_introduction.tex
\noindent
Smeared spectral densities constitute a phenomenologically important class of observables in quantum field theories, especially quantum chromodynamics (QCD).
Mathematically, the underlying spectral functions are
singular tempered distributions, so it is natural to treat a generic spectral density $\rho$ as a linear functional acting on test kernels.
For a given smearing kernel $\kappa$, the corresponding smeared observable will be denoted by
\begin{equation}\label{eq:rho_kappa}
    \rho[\kappa] = \int_0^\infty d\omega \, \rho(\omega) \kappa(\omega) \,.
\end{equation}
The characterization of spectral functions through their smeared versions is appealing for several reasons.

First, phenomenological applications often dictate a particular form for the smearing kernel $\kappa$, e.g., via the kinematics of an inclusive decay.
Relevant examples include hadronic contributions to the muon anomalous magnetic moment~\cite{Aliberti:2025beg} and inclusive semileptonic decays~\cite{Bailas:2020qmv,Gambino:2020crt,Evangelista:2023fmt,ExtendedTwistedMass:2024myu,DeSantis:2026iuu}.

Second, smearing provides a rigorous bridge between different determinations of spectral densities, either theoretical or experimental~\cite{Juttner:2026fui}.
For instance, weighted spectral integrals are used to relate experimentally determined hadronic $\tau$ decay spectral functions to the isovector contribution to $e^+e^- \to \text{hadrons}$ cross sections~\cite{Davier:2005xq}.
Additionally, smearing is not only possible but often essential to provide a meaningful comparison between theory and experiments~\cite{Poggio:1975af}.
This has been particularly relevant to describe inclusive hadronic scattering amplitudes and decay rates from first principles lattice field theories~\cite{Luscher:1988sd,Maiani:1990ca,Barata:1990rn,Hansen:2017mnd,Bulava:2019kbi,Bruno:2020kyl,Patella:2024cto}, a prime example being the determination of the $R$ ratio~\cite{Cabibbo:1961sz,Cabibbo:1970mh} from lattice QCD (LQCD)~\cite{ExtendedTwistedMassCollaborationETMC:2022sta}.

Finally, and closely related to the previous points, it is desirable to compute smeared spectral functions non-perturbatively directly from the underlying theory of hadronic physics, QCD, and in particular from systematically improvable LQCD. 
Since lattice field theory calculations are performed in a finite spatial volume, the associated finite-volume spectral density becomes a weighted sum of Dirac delta functions.
Aside from being a technical necessity to extract meaningful observables from such a distribution, evaluating the functional $\rho[\kappa]$ with a smooth smearing kernel additionally mitigates finite-volume effects~\cite{Bresciani:2026kjv,Hansen:2017mnd,Bulava:2021fre}, which shift both the amplitudes and the energies of the discrete spectrum.
In this way, smearing supplies the bridge between finite-volume quantities computed using LQCD, infinite-volume spectral densities, and experimentally accessible observables, beyond the range of multiparticle states and energy regions currently accessible through quantization conditions~\cite{Luscher:1985dn,Luscher:1986pf,Luscher:1990ux,Rummukainen:1995vs,Lellouch:2000pv,Kim:2005gf,Hansen:2014eka,Briceno:2017max,Hansen:2019nir}.

Numerical evaluation of the path integral using LQCD does not provide direct access to arbitrarily smeared spectral functions $\rho[\kappa]$, but rather to noisy and discrete samples of Euclidean time correlation functions.
The spectral representation shows that zero-temperature Euclidean-time two-point correlation functions of Hermitian operators can themselves be viewed as smeared spectral observables\footnote{The present discussion focuses on zero-temperature two-point functions for the sake of simplicity.
The finite-temperature case can be tackled similarly by suitably redefining $\phi_t(\omega)$.
Generalization to higher-point functions is also possible but exceeds the scope of the present work.}:
\begin{align}
    \label{eq:inverse_problem}
    C(t) = \int_0^\infty d\omega \, \rho(\omega) e^{-\omega |t|} \equiv \rho[\phi_t] \,.
\end{align}
In this case, the smearing function is given by the Laplace kernel, $\phi_t(\omega) = e^{-\omega |t|}$.
Extracting a \emph{generic} target observable $\rho[\kappa]$ directly from $\rho[\phi_t]$ constitutes a well-known inverse problem. For realistic lattice applications, this problem is additionally rendered ill-posed by the noisy, finite and discrete nature of Monte Carlo data for Euclidean-time lattice correlators.

Any practical solution to the inverse problem of extracting $\rho(\omega)$ from \cref{eq:inverse_problem} must grapple with two conceptually distinct sources of uncertainty.
The first arises even when input points are exact: a finite, discrete set of input data does not uniquely determine the spectral function $\rho(\omega)$.
More precisely, when the problem is formulated as a truncated moment problem~\cite{Abbott:2025snz}, such data can at most bound spectral functions smeared with a Cauchy kernel
to lie within a calculable finite interval at each point~\cite{Kovalishina1984}.
Recent extensions have made similar bounds available for other choices of smearing kernels~\cite{Lawrence:2024hjm,Abbott:2026wdw,Mutzel:2026vyw}.
These bounds quantify the unavoidable epistemic uncertainty intrinsic to reconstructing a smeared, and hence regulated, spectral function from only finitely many data.
The second source of uncertainty is statistical and arises from the Monte Carlo estimation of the input data.
Regardless of the method used to solve the spectral inverse problem, the reliable construction of complete error budgets is widely recognized as crucial for phenomenological applications~\cite{Bulava:2021fre,Bruno:2023bue,DiCarlo:2025mnm,Candido:2024hjt,DelDebbio:2022qgu,Gambino:2022dvu,Barone:2023tbl,Kellermann:2025pzt,Kellermann:2026sgp,DeSantis:2025yfm,DeSantis:2025qbb,ExtendedTwistedMassCollaborationETMC:2022sta,Evangelista:2023fmt,ExtendedTwistedMass:2024myu,Bennett:2024cqv,TELOS:2025ash,Buzzicotti:2023qdv}.

The bulk of the present work begins one step later. 
The idea of the kernel-transformation problem is to use a family of energy-smeared observables as input to compute the same spectral function smeared with another kernel.
Different smearing kernels possess distinct theoretical and practical virtues.
For instance, Gaussian kernels have been argued to be useful for approaching the unsmeared limit~\cite{Bulava:2021fre,Bulava:2023mjc}.
Meanwhile, the analytic structure of Cauchy and regulated principle-value kernels is closely related to the $i\varepsilon$ prescription and Lehmann--Symanzik--Zimmermann reduction formalism for scattering amplitudes~\cite{Bulava:2019kbi}.
Other kernels of interest include, but are not limited to, kinematically motivated kernels.
Kernel transformations may then be leveraged to define a convenient intermediate class of observables from which bounds on different regulated extractions of $\rho(\omega)$ from \cref{eq:inverse_problem} can be more easily constructed.

In general, the input data for kernel transformations take the form of spectral integrals
\begin{equation}\label{eq:rho_alpha}
    \rho[\phi_\alpha] = \int_0^\infty d\omega \, \rho(\omega) \phi_\alpha(\omega)
\end{equation}
over a given family of input kernels, denoted as $\{\phi_\alpha\}$, where $\alpha \in A$ is a discrete or continuous label in some given set $A$.
In the applications below, $\alpha$ can represent an energy center, a smearing width or a Euclidean time as in \cref{eq:inverse_problem}.
The reconstruction of the target observable $\rho[\kappa]$ from the input data $\rho[\phi_\alpha]$ defines the kernel-transformation problem, i.e., the problem of constructing a map between two families of kernels.

The central novel contributions of the present work are as follows.
First, we develop the formalism of linear kernel-to-kernel transformations, establishing conditions under which these maps exist and converge without requiring arbitrary regularization, avoiding distortion of the target smeared observable.
Second, we provide explicit analytic expressions for several transformations between common smearing kernels, with the Cauchy-to-Gaussian map in \cref{eq:Cauchy_to_Gaussian_exact} serving as a primary worked example and motivation.
Third, we show how rigorous pointwise bounds on an input smeared spectral function can be propagated directly through a transformation using a sign-splitting argument inspired by Riesz-Kantorovich operator theory, leading to \cref{eq:lower_bound_Krho,eq:upper_bound_Krho}. 
Fourth, we formulate the propagation of bounds in a general kernel transformation as a convex-optimization problem, imposing spectral positivity and leading to certified upper and lower bounds in \cref{eq:rho_kappa_pm_rig}.
The two methods naturally extend to the case of statistical error propagation.
Finally, whenever an approximation is introduced to render a kernel transformation practically tractable, we show how to certify the associated systematic error from the known input data in \cref{eq:bound_rho_kappa_from_data}, leading to the optimal certified upper and lower bounds on $\rho[\kappa]$ in \cref{eq:bounds_regulator_T,eq:rho_kappa_pm_rig_RK}.

The remainder of the paper is structured as follows.
\cref{sec:formalism} develops the formalism of exact and regulated kernel transformations, with particular emphasis on analytically solvable cases and on the obstructions that force regularization.
\cref{sec:bounds} describes the propagation of rigorous bounds under such transformations, including both Riesz-Kantorovich bounds and positivity-preserving optimized bounds, and discusses the optimization of input smearing and regulator choices.
\cref{sec:numerics} provides numerical demonstrations of the formalism for optimized pointwise error propagation in an exact Cauchy-to-Gaussian transformation, a regulated Cauchy-to-Cauchy sharpening problem, and a Gaussian-to-Cauchy width-mixture transformation. 
Statistical noise is included in a final numerical example, where Euclidean-time correlator data is transformed to Cauchy- and Gaussian-smeared spectral functions.
Several Appendixes provide supporting derivations and technical details.

Code for the numeric examples in Sec.~\ref{sec:numerics} is available at Ref.~\cite{code_github}.

%% file: sec_kernel_transformations.tex
\noindent
The preceding discussion of kernel transformations as maps between families of kernels suggests a natural strategy to compute $\rho[\kappa]$ from $\rho[\phi_\alpha]$.
Rather than targeting generically ill-defined direct reconstruction of $\rho(\omega)$, one can leverage linearity to express the target kernel $\kappa(\omega)$ as a linear combination of the input kernel family $\{\phi_\alpha\}$.
The aim is to define a set of ``transition kernels'' $K^{\kappa\leftarrow\phi}(\alpha)$ satisfying
\begin{equation}\label{eq:kernel_expansion}
    \kappa(\omega) = \int_A d\alpha \, K^{\kappa\leftarrow\phi}(\alpha) \, \phi_\alpha(\omega) \,.
\end{equation}
Whenever this representation is convergent, and the interchange of integrations over $\omega$ and $\alpha$ is justified, linearity of the functional $\rho[\kappa]$ yields a direct map from the input data to the target observable,
\begin{equation}\label{eq:kernel_transformation}
    \rho[\kappa]
    =
    \int_A d\alpha \, K^{\kappa\leftarrow\phi}(\alpha) \, \rho[\phi_\alpha] \,.
\end{equation}
When such a map can be constructed, it provides the desired kernel transformation from the input data to the target observable.

The validity of the formal construction in \cref{eq:kernel_transformation} depends strongly on the analytic properties of the kernels involved.
An elegant illustration of the success of this approach is provided by the time-momentum representation of the hadronic contributions to the muon anomalous magnetic moment, $a_\mu^{\rm{HVP}}$~\cite{Bernecker:2011gh}.
More generally, for
$\phi_t(\omega) = e^{-\omega t}$ in \cref{eq:rho_alpha}, the feasibility of kernel transformations is known to be closely related to the analytic properties of the target smearing kernel.
In particular, for LQCD applications, where the existence of a mass gap $\omega_0>0$ restricts the support of $\rho(\omega)$ to $\omega > \omega_0$, transition kernels from $\phi_t=e^{-\omega t}$ can be explicitly defined via analytic continuation only as long as any nonanalytic structure of the target kernel $\kappa(\omega)$ is confined to points $\omega_i \in \mathbb{C}$ below the mass gap, i.e., $\real \omega_i < \omega_0$.
In this case, the Bromwich integral defining the inverse Laplace transform can be evaluated on a deformed contour, which is accessible directly from the Euclidean time domain.
This situation is indeed the case for $a_\mu^{\rm{HVP}}$.
A similar set of restrictions will be shown to hold for generic kernel transformations.

The construction of generic kernel transformations hinges on three logically distinct requirements:
\begin{enumerate}
    \item the input family of kernels must be complete on the space of target kernel(s),
    \item the transition kernel must exist as in \cref{eq:kernel_transformation} without the need for regularization to tame uncontrolled divergences, and
    \item the interchange of the $\omega$ and $\alpha$ integrations must be justified.
\end{enumerate}
Failure of any of these requirements prevents the unambiguous construction of an exact kernel transformation.

The remainder of this Section develops a framework, based largely on standard techniques from Fourier analysis, to construct kernel transformations when they exist and to understand the analytic obstructions when they do not.
\Cref{subsec:general-realization} formalizes the approach and establishes notation.
\Cref{subsec:analytic_kernel_transformations} treats the core examples of Cauchy and Gaussian smearing kernels used throughout the paper.
\Cref{subsec:phenomenological-targets} records how the same criteria apply to finite-window and principal-value targets.
\Cref{subsec:formalism-regulated} discusses the construction of regulated transition kernels and the possibility to estimate the associated bias.

\subsection{General realization of kernel transformations}\label{subsec:general-realization}
\noindent
Many smearing kernels display a dependence on continuous parameters beyond the integration variable (cf. \cref{eq:rho_kappa,eq:rho_alpha}).
Common examples include Cauchy and Gaussian smearing kernels
\begin{align} \label{eq:delta_cauchy}
    K^{\ctt}_\varepsilon(\omega,\omega')
    &=
    \frac{\varepsilon/\pi}{(\omega-\omega')^2+\varepsilon^2}
    \equiv
    \delta^{\ctt}_\varepsilon(\omega -\omega') \,,\\
    \label{eq:delta_gaussian}
    K^{\gtt}_\sigma(\omega,\omega')
    &=
    \frac{1}{\sqrt{2\pi\sigma^2}}
    e^{-\frac{(\omega-\omega')^2}{2\sigma^2}}
    \equiv \delta^{\gtt}_\sigma(\omega -\omega') \,,
\end{align}
which also depend on the width ($\varepsilon$ and $\sigma$, respectively) and center ($\omega'$) of the smearing.
For such kernels, \cref{eq:rho_kappa} can be rewritten as
\begin{equation}\label{eq:K-generic}
    \rho^{\ktt}_\varepsilon(\omega)
    = \int_0^\infty d\omega' \, K^{\ktt}_\varepsilon(\omega,\omega') \rho(\omega')
    \equiv [ K^{\ktt}_\varepsilon \rho ](\omega) \,,
\end{equation}
where $\mathtt k \in \{ \mathtt{c}, \mathtt{g} \}$ and the shorthand notation
\begin{align}
\rho^{\ktt}_\varepsilon(\omega) \equiv [K^{\ktt}_\varepsilon \rho](\omega) \equiv \rho[\kappa](\omega)
\end{align}
is introduced for convenience.
The additional parameters provide a handle to construct the transition kernels in \cref{eq:kernel_expansion}, by identifying them with the parameters $\alpha$.
For example, focusing on the dependence on $\alpha\equiv\omega$ at fixed $\varepsilon$, the goal is to construct the linear operator
\begin{equation}
    K^{\ktt_2 \leftarrow \ktt_1}_
    {\varepsilon_2 \leftarrow \varepsilon_1}:
    \rho^{\ktt_1}_{\varepsilon_1}(\omega)
    \mapsto
    \rho^{\ktt_2}_{\varepsilon_2}(\omega)
\end{equation}
satisfying \cref{eq:kernel_expansion}, i.e.
\begin{equation}\label{eq:K_k1k2_def}
\begin{aligned}
    \rho^{\ktt_2}_{\varepsilon_2}(\omega') &=
    [K^{\ktt_2 \leftarrow \ktt_1}_{\varepsilon_2 \leftarrow \varepsilon_1}
    \rho^{\ktt_1}_{\varepsilon_1}] (\omega') \\
    & = \int_{\R} d\omega \,
    K^{\ktt_2 \leftarrow \ktt_1}_{\varepsilon_2 \leftarrow \varepsilon_1}(\omega',\omega)
    \rho^{\ktt_1}_{\varepsilon_1}(\omega) \,.
\end{aligned}
\end{equation}
Appendix~\ref{app:normalization} contains additional details relating the normalization of input and target smearing functions with that of the transition kernels.

Postponing momentarily questions of existence and convergence, one formally finds
\begin{align}
    \rho^{\ktt_2}_{\varepsilon_2}(\omega)
    &= [K_{\varepsilon_2}^{\ktt_2} \rho](\omega) 
    \label{eq:target}\\
    &= [
        K_{\varepsilon_2}^{\ktt_2}
        (K_{\varepsilon_1}^{\ktt_1})^{-1}]
        [K_{\varepsilon_1}^{\ktt_1} \rho](\omega)\\
    &= [
        K_{\varepsilon_2}^{\ktt_2}
        (K_{\varepsilon_1}^{\ktt_1})^{-1}]
        \rho^{\ktt_1}_{\varepsilon_1}(\omega)
    \label{eq:target_via_input}
\end{align}
where $(K_{\varepsilon_1}^{\ktt_1})^{-1}$ indicates the formal operator inverse of $K_{\varepsilon_1}^{\ktt_1}$.
The linear map that transforms between kernels is therefore given formally by
\begin{align}\label{eq:kernel_transformation_general}
    K^{\ktt_2 \leftarrow \ktt_1}_
    {\varepsilon_2 \leftarrow \varepsilon_1}(\omega', \omega)
    = \int_{\R} d\omega''\,
        K_{\varepsilon_2}^{\ktt_2}(\omega', \omega'')
        (K_{\varepsilon_1}^{\ktt_1})^{-1}(\omega'', \omega) \,.
\end{align}
In other words, \cref{eq:kernel_expansion} is explicitly realized as
\begin{align}\label{eq:kernel_transformation_general2}
        K_{\varepsilon_2}^{\ktt_2}(\omega, \omega') =
        \int_{\R} d\omega''\,
        K^{\ktt_2 \leftarrow \ktt_1}_{\varepsilon_2 \leftarrow \varepsilon_1}(\omega, \omega'')
        K_{\varepsilon_1}^{\ktt_1}(\omega'', \omega') \,.
\end{align}
Of course, the existence of the inverse and the convergence of the integrals above are essential for any practical use of these expressions.
The key point of this construction is that, although single inverse integral operators might remain analytically intractable and possibly diverge, only the product $K_{\varepsilon_2}^{\kappa_2}(K_{\varepsilon_1}^{\kappa_1})^{-1}$ is actually required.
When the latter converges, exact kernel transformations can be realized without any approximation or regularization (beyond the standard, controlled ones possibly required from a numerical approximation of the exact integrals, as customary e.g. in $a_\mu^\text{HVP}$ applications~\cite{Bernecker:2011gh}).

For a general pair of kernels, computing $K^{\ktt_2 \leftarrow \ktt_1}_{\varepsilon_2 \leftarrow \varepsilon_1}$ may be analytically cumbersome, both for convergent integrals as well as for cases requiring regularization.
In any case, it is envisioned that the transition kernel can be approximated numerically, effectively regulating the problem.
Irrespective of the details of this realization, \cref{eq:target} can be defined via intermediate regularized products in \cref{eq:target_via_input} even when $K^{\ktt_2 \leftarrow \ktt_1}_{\varepsilon_2 \leftarrow \varepsilon_1}$ remains formally inaccessible by itself.

Perhaps simplest strategy is to discretize the energies along a suitable grid, so that the right-hand side of \cref{eq:K-generic} should be understood as a matrix-vector product. 
Other approaches are possible to approximate the expansion of the target on the input kernels in \cref{eq:kernel_transformation_general2}.
For instance, robust approximations with convergence guarantees anchored on the Stone-Weierstrass theorem 
might emerge by minimizing a certain notion of distance (e.g., $L^2$ in Ref.~\cite{Hansen:2019idp}) between the approximate and target kernels.
Given any explicit regularization of the problem, Sec.~\ref{subsec:formalism-regulated} discusses a method to quantify the associated systematic effects. 
This quantification does not depend on the details of the realization of the approximate target smearing kernel and is phrased in terms of quantities which can be directly accessed from the input data only.

\subsection{Gaussian and Cauchy kernel transformations \label{subsec:analytic_kernel_transformations}}
\noindent
For translationally invariant input kernels
\begin{equation}
    K^{\ktt_1}_{\varepsilon_1}(\omega,\omega') = \delta_{\varepsilon_1}^{\ktt_1}(\omega-\omega') \,,
\end{equation}
the kernel transformation map of \cref{eq:kernel_transformation_general2} can be represented analytically in Fourier space.
With the Fourier conjugate of the kernel defined via
\begin{equation}\label{eq:FT}
    \hat\delta_{\varepsilon}^{\ktt}(\tau) \equiv \int_{\R} \frac{d\omega}{2\pi}
    \delta_{\varepsilon}^{\ktt}(\omega)e^{-i\omega \tau} \,,
\end{equation}
the convolution theorem gives the transition kernel\footnote{
    The factor of $2\pi$ in the denominator is a consequence of the convention in \cref{eq:FT}, which implies that
    \begin{gather*}
        f(x) = \int_{\R} dy \, K(x-y) g(y) \,, \\
        \label{eq:FT-convention}
        \hat f(\tau) \equiv \int_{\R} \frac{dx}{2\pi} f(x) e^{-i\tau x}
             = 2\pi \hat K(\tau) \hat g(\tau) \,, \\
        \label{eq:IFT-convention}
        K(x) = \int_{\R} d\tau \, \hat K(\tau) e^{+i\tau x}
             = \int_{\R} \frac{d\tau}{2\pi} \hat f(\tau) \hat g(\tau)^{-1} e^{+i\tau x} \,.
    \end{gather*}
    \Cref{eq:transition_kernel_general} directly comes from these relations, upon identifying $f \mapsto K_{\varepsilon_2}^{\ktt_2}$ and $g \mapsto K_{\varepsilon_1}^{\ktt_1}$.
}
\begin{equation}\label{eq:transition_kernel_general}
    K^{\ktt_2 \leftarrow \ktt_1}_{\varepsilon_2 \leftarrow \varepsilon_1}(\omega', \omega)
    = \int_{\R} \frac{d\tau}{2\pi} \hat K_{\varepsilon_2}^{\ktt_2}(\omega', \tau)
    \hat\delta_{\varepsilon_1}^{\ktt_1}(\tau)^{-1} e^{+i\tau\omega}
\end{equation}
where $\hat K_{\varepsilon_2}^{\ktt_2}(\omega', \tau)$ indicates the Fourier transform of $K_{\varepsilon_2}^{\ktt_2}(\omega', \omega)$ with respect to $\omega$ only.
If $K_{\varepsilon_2}^{\ktt_2}$ also depends on the difference $\omega-\omega'$ only through $K^{\ktt_2}_{\varepsilon_2}(\omega,\omega') = \delta_{\varepsilon_2}^{\ktt_2}(\omega-\omega')$, translational invariance gives
\begin{equation}\label{eq:K_k1k2}
    K^{\ktt_2 \leftarrow \ktt_1}_{\varepsilon_2 \leftarrow \varepsilon_1}(\omega', \omega)
    = \int_{\R} \frac{d\tau}{2\pi} \hat \delta_{\varepsilon_2}^{\ktt_2}(\tau)
    \hat\delta_{\varepsilon_1}^{\ktt_1}(\tau)^{-1} e^{+i\tau(\omega-\omega')}
\end{equation}
which is a function of $\omega'-\omega$, as expected.
For the sake of concreteness and practical interest, consider the Cauchy and Gaussian kernels defined in \cref{eq:delta_cauchy,eq:delta_gaussian}.
Their Fourier conjugates are
\begin{align}
    \label{eq:fourier_transform_gaussian}
    \hat{\delta}_\sigma^{\gtt}(\tau)
    &= \int_{\R} \frac{d \omega}{2\pi}
    \delta_\sigma^{\gtt}(\omega)e^{-i\omega \tau}
    = \frac{e^{-\tfrac{1}{2}\tau^2 \sigma^2}}{2\pi} \,, \\
    \label{eq:fourier_transform_cauchy}
    \hat{\delta}_\varepsilon^{\ctt}(\tau)
    &= \int_{\R} \frac{d \omega}{2\pi}
    \delta_\varepsilon^{\ctt}(\omega)e^{-i \omega \tau}
    = \frac{e^{-\varepsilon|\tau|}}{2\pi} \,.
\end{align}
The next three examples form the core exact transformations used to organize the later discussion.
The Gaussian-to-Gaussian and Cauchy-to-Cauchy cases show the familiar widening semigroups, while the Cauchy-to-Gaussian case provides the main nontrivial exact bridge used below for bound propagation and numerical tests.
The discussion then turns to transformations that fail to converge in the fixed-width Fourier representation of \cref{eq:K_k1k2}, as well as to a different transformation realized by a linear combination of kernels centered around the same energy but with different smearing widths.

\paragraph{Gaussian to Gaussian.}
The transition kernel between two Gaussians is given via \cref{eq:K_k1k2} by
\begin{align}
    K_{\sigma_2 \leftarrow \sigma_1}^{\gtt \leftarrow \gtt}(\omega'-\omega)
    &= \int_{\R} \frac{d\tau}{2\pi}
    \hat{\delta}_{\sigma_2}^{\gtt}(\tau)
    \hat{\delta}_{\sigma_1}^{\gtt}(\tau)^{-1} e^{+i \tau (\omega'-\omega)}\\
    &= \int_{\R} \frac{d\tau}{2\pi}
    e^{-\tfrac{1}{2}\tau^2 (\sigma_2^2 - \sigma_1^2)}
    e^{+i \tau (\omega'-\omega)} \\
    \label{eq:g2g}
    &= \delta^{\gtt}_{\sqrt{\sigma_2^2-\sigma_1^2}}(\omega'-\omega) \,.
\end{align}
The integral in the second line is absolutely convergent for $\sigma_2 \geq \sigma_1$ only.
\cref{eq:g2g} defines the well-known semigroup property of Gaussian kernels under convolution.
Unregulated deconvolution ($\sigma_2 < \sigma_1$) is impermissible, while additional smearing ($\sigma_2 > \sigma_1$) presents no obstruction.
Unsurprisingly, Gaussian data can be transformed exactly to a broader Gaussian resolution, but not sharpened by this unregulated map.

\paragraph{Cauchy to Cauchy.}
For the case of two Cauchy kernels, \cref{eq:K_k1k2} becomes
\begin{align}
    K_{\varepsilon_2 \leftarrow \varepsilon_1}^{\ctt \leftarrow \ctt}(\omega'-\omega)
    &= \int_{\R} \frac{d\tau}{2\pi}
    \hat{\delta}_{\varepsilon_2}^{\ctt}(\tau)
    \hat{\delta}_{\varepsilon_1}^{\ctt}(\tau)^{-1} e^{+i \tau (\omega'-\omega)}\\
    &= \int_{\R} \frac{d\tau}{2\pi}
    e^{-(\varepsilon_2 - \varepsilon_1) |\tau|}
    e^{+i \tau (\omega'-\omega)} \\
    \label{eq:c2c}
    &= \delta^{\ctt}_{\varepsilon_2-\varepsilon_1}(\omega'-\omega) \,,
\end{align}
where the integral in the second line is absolutely convergent for $\varepsilon_2 \geq \varepsilon_1$ only.
\Cref{eq:c2c} defines the semigroup property of Cauchy kernels under convolution.
As with the preceding Gaussian-to-Gaussian example, unregulated deconvolution is not defined, while additional smearing proceeds without problem.
As discussed below, this fact can be interpreted in terms of the analytic structure of the Cauchy kernel in the complex plane.
Thus, Cauchy data can also be transformed exactly to a broader Cauchy resolution, while sharpening requires a regulated treatment.

\paragraph{Cauchy to Gaussian.}
For the reconstruction of a Gaussian target smearing from a Cauchy input, the transition kernel from \cref{eq:K_k1k2} takes the form
\begin{align}
    K_{\sigma \leftarrow \varepsilon}^{\gtt \leftarrow \ctt}(\omega', \omega)
    &= \int_{\R} \frac{d\tau}{2\pi} \,
    \hat{\delta}_\sigma^{\gtt}(\tau)
    \hat{\delta}_\varepsilon^{\ctt}(\tau)^{-1} e^{+i \tau (\omega'-\omega)}\\
    \label{eq:c-to-g2}
    &= \int_{\R} \frac{d\tau}{2\pi} \,
    e^{-\tfrac{1}{2}\tau^2 \sigma^2}
    e^{+\varepsilon |\tau|}
    e^{+i \tau (\omega'-\omega)}
    \\
    \label{eq:Cauchy_to_Gaussian_exact}
    &= \frac1{\sqrt{2\pi\sigma^2}} \mathrm{Re} \left\{ e^{z^2} [1+\erf(z)] \right\} \,,
\end{align}
where $z \equiv \left(\varepsilon + i(\omega'-\omega)\right)/\sqrt{2\sigma^2}$.
\begin{figure}[t!]
    \centering
    \includegraphics[width=0.99\linewidth]{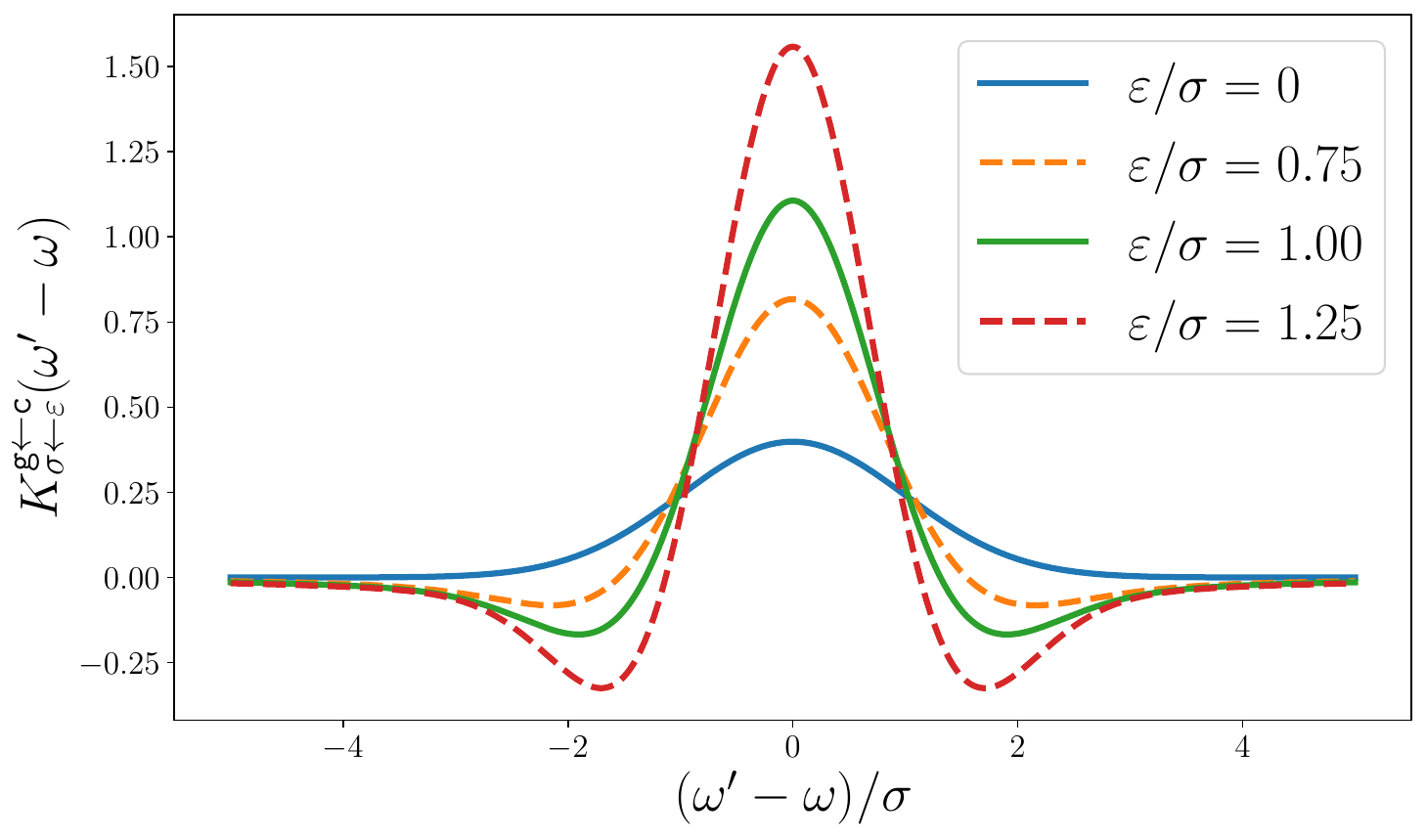}
\caption{Transition kernel $K_{\sigma \leftarrow \varepsilon}^{\gtt \leftarrow \ctt}(\omega' - \omega)$ moving from a Cauchy kernel of width $\varepsilon$ to a Gaussian kernel with fixed width $\sigma$.
The magnitude of oscillations increases dramatically when $\varepsilon\gg\sigma$, while the $\varepsilon\to0$ limit reproduces the target Gaussian, $\delta_\sigma^{\gtt}$. }
    \label{fig:cauchy_to_gaussian_kernel}
\end{figure}
The smeared spectral functions are related via \cref{eq:K_k1k2_def}, which reads
\begin{align}
\rho_{\sigma}^{\mathtt g}(\omega) =
    \int d\omega' K_{\sigma \leftarrow \varepsilon}^{\gtt \leftarrow \ctt}(\omega'-\omega)
    \rho_\varepsilon^{\ctt}(\omega') \,.
\end{align}
Interestingly, the Cauchy-to-Gaussian transition kernel is exactly defined without any regularization for all positive $\varepsilon$ and $\sigma$. 
\Cref{fig:cauchy_to_gaussian_kernel} illustrates the behavior of the transition kernel $K_{\sigma \leftarrow \varepsilon}^{\gtt \leftarrow \ctt}$ for a few representative choices of smearing widths.
As expected, $K_{\sigma \leftarrow \varepsilon}^{\gtt \leftarrow \ctt}$ reduces to the Gaussian kernel when $\varepsilon\to0$.
This exact Cauchy-to-Gaussian map is the principal example used in Secs.~\ref{sec:bounds} and~\ref{sec:numerics} to illustrate the propagation of rigorous bounds through kernel transformations.

\paragraph{Gaussian to Cauchy.}
The reverse transformation, from Gaussian-smeared data at one fixed width to a specified Cauchy target, fails the Fourier-ratio convergence criterion in \cref{eq:K_k1k2}.
In this setting only the center of the Gaussian is varied, and the desired fixed-width transition would require
\begin{align}
    K_{\varepsilon \leftarrow \sigma}^{\ctt \leftarrow \gtt}(\omega', \omega)
    &\stackrel{?}{=} \int_{\R} \frac{d\tau}{2\pi} \,
    \hat{\delta}_\varepsilon^{\ctt}(\tau)
    \hat{\delta}_\sigma^{\gtt}(\tau)^{-1} e^{+i \tau (\omega'-\omega)}\\
    \label{eq:g2c}
    &= \int_{\R} \frac{d\tau}{2\pi} \,
    e^{+\tfrac{1}{2}\tau^2 \sigma^2}
    e^{-\varepsilon |\tau|}
    e^{+i \tau (\omega'-\omega)}\\
    &= \infty \,.
\end{align}
In other words, the transformation from a fixed Gaussian width to a specified Cauchy kernel via \cref{eq:K_k1k2_def} must always be accompanied by some kind of regularization, as described in Sec.~\ref{subsec:formalism-regulated}.

\begin{figure}[t!]
    \centering
    \includegraphics[width=0.99\linewidth]{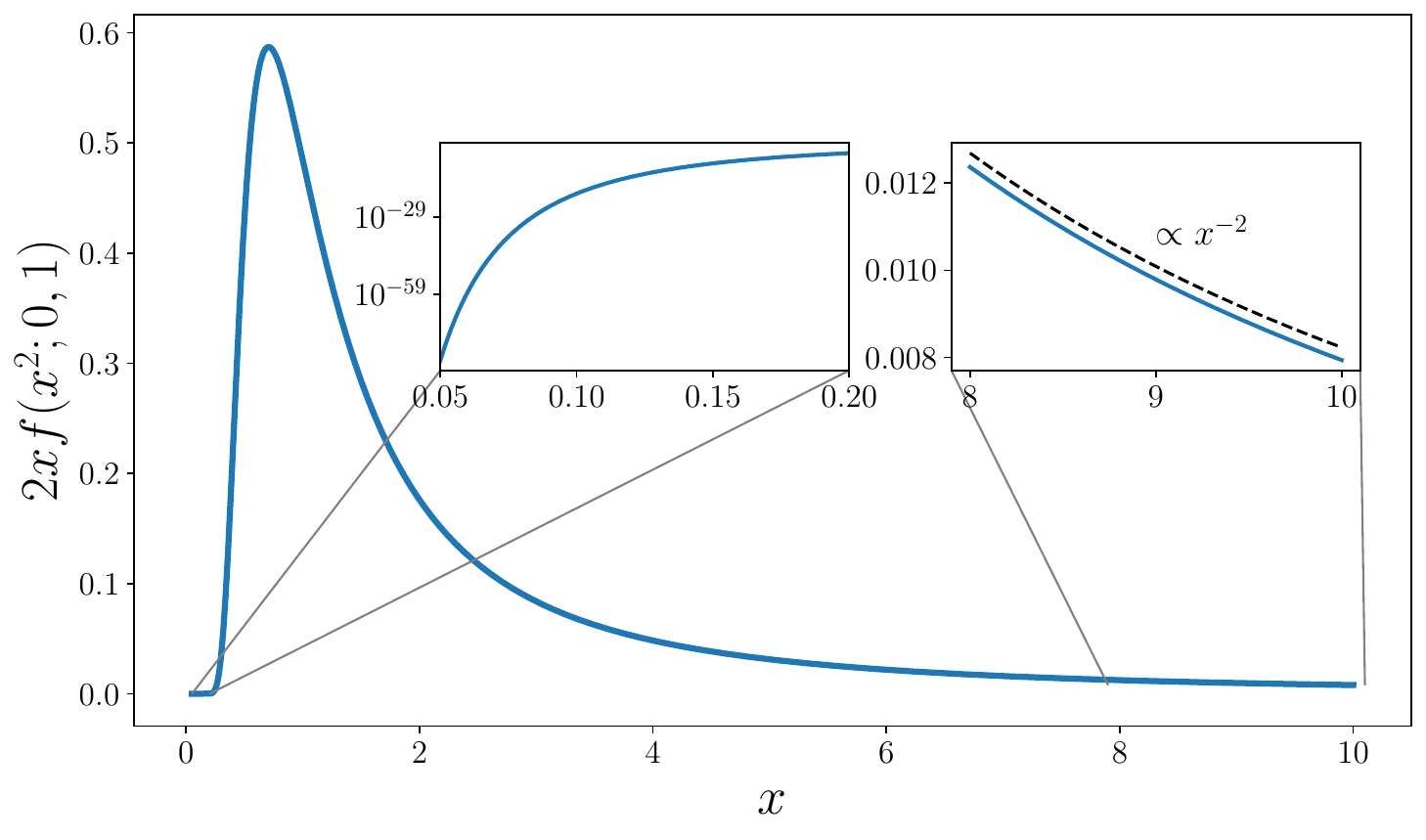}
\caption{Transition kernel for a Gaussian-to-Cauchy width mixture, targeting a Cauchy kernel of width $\varepsilon$ from a family of Gaussians of variable width $\sigma$.
The kernel is shown as a function of $x=\sigma/\varepsilon$.}
    \label{fig:gaussian_to_cauchy_kernel}
\end{figure}

However, this obstruction does not rule out every Gaussian-to-Cauchy construction.
If, instead, one has access to Gaussian-smeared observables over a continuum of widths at a fixed center, a Cauchy-smeared observable at that same center can be formed by a linear mixture of different widths.
The starting point is the elementary integral
\begin{equation}
    \frac1{x} = \int_0^\infty dy \, e^{-yx} \,.
\end{equation}
Taking the denominator on the left-hand side to be that of the Cauchy kernel, $x=(\omega-\omega')^2+\varepsilon^2$, and applying the change of variable $y=1/\sigma^2$ yields the representation
\begin{equation}\label{eq:Levy}
    \delta_\varepsilon^{\ctt}(\omega-\omega') =
    \int_0^\infty d\sigma \, 2\sigma f(\sigma^2;0,\varepsilon^2) \delta_{\sigma}^{\gtt}(\omega-\omega') \,,
\end{equation}
where
\begin{equation}
    f(x;\mu,c) \equiv \sqrt{\frac{c}{2\pi}} \frac{e^{-\frac{c}{2(x-\mu)}}}{(x-\mu)^{3/2}}
\end{equation}
is the probability distribution function of the L\'evy distribution over $x\geq\mu$, with location and scale parameters $\mu$ and $c$, respectively.
\Cref{fig:gaussian_to_cauchy_kernel} shows this kernel.
Integrating out both sides of \cref{eq:Levy} against $\rho(\omega')$ gives
\begin{equation}\label{eq:rho_Levy}
    \rho_\varepsilon^{\ctt}(\omega) = \int_0^\infty d\sigma \, 2\sigma
    f(\sigma^2;0,\varepsilon^2) \rho_{\sigma}^{\gtt}(\omega) \,.
\end{equation}
In other words, a Cauchy-smeared spectral function can be recovered from a linear combination of Gaussian-smeared spectral functions with different widths.\footnote{\label{foot:Levy}
    Changing integration variable in \cref{eq:Levy} from $\sigma$ to the dimensionless quantity $x=\sigma/\varepsilon$ yields
    \begin{equation*}\label{eq:Levy_x}
        \delta_\varepsilon^{\ctt}(\omega,\omega') = \int_0^\infty dx \, 2x
        f(x^2;0,1) \delta_{x\varepsilon}^{\gtt}(\omega,\omega') \,.
    \end{equation*}
    The variable $x$ can be interpreted as the fraction of Gaussian smearing width with respect to the target Cauchy smearing width.
    The corresponding smeared spectral function is then recovered as in \cref{eq:rho_Levy} through
    \begin{equation*}
        \rho_\varepsilon^{\ctt}(\omega) = \int_0^\infty dx \, 2x
        f(x^2;0,1) \rho_{x\varepsilon}^{\gtt}(\omega) \,.
    \end{equation*}
}
This does not contradict the fixed-width failure above: the input data here are the family $\{\rho_\sigma^{\gtt}(\omega)\}_{\sigma>0}$ at the chosen center, rather than a single Gaussian width as the center varies.
Note that contributions from small values of $\sigma$ are exponentially suppressed by the L\'evy function's zero of infinite order at $\sigma=0$.
This behavior may be beneficial for practical reconstructions, since $\rho_\sigma^{\gtt}$ is typically the most difficult to obtain in that region, with the uncertainties increasing sharply for $\sigma\to 0$.

The analogous reverse width mixture cannot be defined unambiguously, as it would require a function $g(\varepsilon)$ satisfying
\begin{equation}
    \delta_\sigma^{\gtt}(\omega) = \int_0^\infty d\varepsilon \, g(\varepsilon) \delta_{\varepsilon}^{\ctt}(\omega) \,.
\end{equation}
Taking the Fourier transform of both sides, using \cref{eq:fourier_transform_gaussian,eq:fourier_transform_cauchy} reveals $g(\varepsilon)$ to be the inverse Laplace transform of $e^{-\tfrac{1}{2}\sigma^2 \tau^2}$,
\begin{equation}
    e^{-\tfrac{1}{2}\sigma^2 \tau^2} = \int_0^\infty d\varepsilon \, g(\varepsilon) e^{-\varepsilon|\tau|} \,,
\end{equation}
which in turn does not exist, even as a highly singular or tempered distribution.

\paragraph{Cauchy to a generic kernel.}
The preceding examples involving Cauchy kernels can be understood as special cases of a general analytic criterion.
Consider the transformation from a Cauchy kernel to an arbitrary, translationally invariant kernel $\kappa(\omega',\omega) = \delta^{\ktt}_\sigma(\omega'-\omega)$ with Fourier conjugate $\hat{\delta}^{\ktt}_\sigma(\tau)$.
The question of existence again hinges on the convergence of \cref{eq:K_k1k2} in the form
\begin{align}\label{eq:K_cauchy_to_general}
    K_{\sigma \leftarrow \varepsilon}^{\ktt \leftarrow \ctt}(\omega' - \omega)
    &= \int_{\R} \frac{d\tau}{2\pi}
    \hat{\delta}^{\ktt}_\sigma(\tau)
    e^{+\varepsilon|\tau|} e^{+i \tau (\omega'-\omega)} \,.
\end{align}
A practical sufficient condition can be stated in terms of the analytic structure of $\delta^\ktt_\sigma(z)$, considered as a function of $z\in \C$.
First, we note that the exponential decay of the Fourier transform of the Cauchy kernel is given by the imaginary part $\varepsilon$ of its poles. Generalizing this observation, \cref{eq:K_cauchy_to_general} is guaranteed to converge as long as $\delta^\ktt_\sigma(z)$ is 
a meromorphic function whose poles $z_i$ lie farther from the real line than the input Cauchy width, i.e., satisfying $\min_i \vert \imag z_i \vert > \varepsilon$.
This result is a consequence of the classical Paley-Wiener theorem~\cite{Paley:1935} (see also Theorem IX.13 of Ref.~\cite{Reed:1975uy}).

The preceding results may also be interpreted in light of this theorem.
For example, the success of the Cauchy-to-Cauchy smearing for $\varepsilon_2 > \varepsilon_1$ in \cref{eq:c2c} is an immediate corollary.
The failure of fixed-width Gaussian-to-Cauchy smearing in \cref{eq:g2c} is the $\varepsilon\to\infty$ limiting case of this general result, ultimately traceable to the Gaussian kernel's essential singularity at complex infinity.
Thus, Cauchy-smeared data of width $\varepsilon$ can be mapped exactly to target kernels whose nearest complex singularities are separated from the real axis by more than $\varepsilon$.

\subsection{Phenomenologically motivated target kernels}\label{subsec:phenomenological-targets}
\noindent
The same convergence logic applies to target kernels that are not in the classes of Gaussian or Cauchy test smearings considered above.
The following examples indicate how convergence requirements emerge from the Fourier-space damping of the target kernel.

\paragraph{Finite-interval observables.}
Many physical observables can be expressed as weighted integrals of the spectral function over a finite interval of the real line
\begin{align}\label{eq:integral_finite_domain}
    \Gamma = \int_{\omega_0}^{\omega_1} d\omega \, R(\omega) \rho(\omega)
    = \int_{\R} d\omega \, R(\omega) W(\omega; \omega_0, \omega_1)
    \rho(\omega) \,,
\end{align}
where $R(\omega)$ is a known rational function and $W(\omega;\omega_0, \omega_1)$ is a rectangular window with support on $[\omega_0, \omega_1]$, i.e.,
\begin{align}\label{eq:hard_window}
    W(\omega; \omega_0, \omega_1)
    = \vartheta(\omega-\omega_0) \vartheta(\omega_1 - \omega) \,,
\end{align}
where $\vartheta(\cdot)$ is the Heaviside step function.
In physical terms, the lower integration bound $\omega_0$ is often related to the presence of a mass gap in the theory, below which the spectral function does not have support.
The upper bound $\omega_1$ instead typically comes from a phase-space threshold or cutoff, e.g., the mass of a heavy decaying particle.
Kinematic considerations usually dictate the functional form of the rational function $R(\omega)$.

The abrupt bounds of the window function in \cref{eq:hard_window} are often smoothed in applications~\cite{Gambino:2020crt,Gambino:2022dvu,Evangelista:2023fmt}.
Let us introduce the following smeared version of the function $R(\omega) W(\omega;\omega_0, \omega_1)$
\begin{align}\label{eq:smeared_window}
    RW_\sigma^{\ktt_2}(\omega; \omega_0, \omega_1) \equiv \int d\omega' \,
    \delta^{\ktt_2}_\sigma(\omega'-\omega) R(\omega') W(\omega'; \omega_0, \omega_1) \,,
\end{align}
with $\delta^{\ktt_2}_\sigma(\omega) \to \delta(\omega)$ as $\sigma\to0$.
A possible choice for this purpose is either kernel in \cref{eq:delta_gaussian,eq:delta_cauchy}, but other alternatives can be considered.
\Cref{eq:smeared_window} induces a smoothed version of the integral
\begin{align}
    \Gamma^{\ktt_2}_\sigma
    &= \int d\omega \, RW^{\ktt_2}_\sigma(\omega) \rho(\omega) \\
    &= \int d\omega' \, \int_{\R} d\tau \, \widehat{RW^{\ktt_2}_\sigma}(\tau; \omega_0, \omega_1)
    e^{i\tau\omega'} \rho(\omega') \,.
\end{align}
To rewrite the integral above in terms of the (input) spectral function smeared with $\delta_\varepsilon^{\ktt_1}(\omega)$, it is useful to leverage the identity
\begin{equation}
    e^{i\tau\omega'} = \frac{1}{2\pi \hat\delta_\varepsilon^{\ktt_1}(\tau)}
    \int d\omega \, \delta_\varepsilon^{\ktt_1}(\omega-\omega') e^{i\tau\omega} \,,
\end{equation}
which holds for all $\varepsilon>0$.
Exchanging the order of integrations gives
\begin{align}
    \label{eq:linear_map_to_finite_integral_step1}
    \Gamma^{\ktt_2}_\sigma
    &= \int d\omega \, \left[ \int \frac{d\tau}{2\pi}
    \frac{\widehat{RW^{\ktt_2}_\sigma}(\tau; \omega_0, \omega_1)}{\hat{\delta}^{\ktt_1}_\varepsilon(\tau)} e^{i \tau \omega} \right]
    \rho^{\ktt_1}_\varepsilon(\omega) \\
    \label{eq:linear_map_to_finite_integral}
    &= \int d\omega \, \left[ \int d\tau\, \widehat{R W}(\tau; \omega_0, \omega_1)
    \frac{\hat{\delta}^{\ktt_2}_\sigma(\tau)} {\hat{\delta}^{\ktt_1}_\varepsilon(\tau)}
    e^{i \tau \omega} \right] \rho_\varepsilon^{\ktt_1}(\omega) \,,
\end{align}
where the convolution theorem has been used in the second identity, and $\widehat{RW}$ denotes the Fourier transform of the rational function on the window, $R(\omega) W(\omega;\omega_0,\omega_1)$.
Whenever it converges, the final line defines the transition kernel mapping the smeared spectral function $\rho_\varepsilon^{\ktt_1}(\omega)$ to a smeared version of \cref{eq:integral_finite_domain} satisfying $\Gamma^{\ktt_2}_\sigma \to \Gamma$ as $\sigma \to 0$.

In order for \cref{eq:linear_map_to_finite_integral} to provide a well-defined transition kernel, the decay of $\widehat{RW}(\tau; \omega_0, \omega_1) \hat\delta_\sigma^{\ktt_2}(\tau)$ must compensate the growth of $\hat\delta_\varepsilon^{\ktt_1}(\tau)^{-1}$.
The sharp window by itself gives only power-law Fourier damping.
This can be seen directly in the special case $R(\omega)=1$, for which
\begin{align}
    \widehat{RW}(\tau;\omega_0, \omega_1)
    &=
    \widehat{W}(\tau; \omega_0, \omega_1)
    =
    \int_{\omega_0}^{\omega_1} \frac{d\omega}{2\pi}  e^{-i \omega \tau} \\
    &= \frac{i}{2\pi\tau}\left( e^{-i\omega_1 \tau} - e^{-i \omega_0 \tau}\right) \,,
\end{align}
which is finite at $\tau=0$ and decays as $1/\tau$ as $|\tau|\to \infty$.
The same asymptotic behavior holds for a nontrivial rational function $R(\omega)$, provided it is analytic in the interval $[\omega_0, \omega_1]$.
The power-law scaling follows from integration by parts:
\begin{equation}
\begin{aligned}
    \widehat{RW}(\tau; \omega_0, \omega_1)
    &= \frac{i}{2\pi\tau} \bigg\{
        [ R(\omega_1) e^{-i\omega_1 \tau} - R(\omega_0) e^{-i\omega_0 \tau} ]  \\
        & \quad\quad\quad\quad - \int_{\omega_0}^{\omega_1} d\omega \, R'(\omega) e^{-i \omega \tau}
    \bigg\} \,.
\end{aligned}
\end{equation}
By the Riemann-Lebesgue lemma and analyticity of $R$ in the interval $[\omega_0, \omega_1]$, the remaining integral is bounded and tends to zero as $|\tau|\to\infty$.
Thus the sharp-window contribution is $O(1/|\tau|)$.

The convergence of \cref{eq:linear_map_to_finite_integral} is therefore guaranteed, provided that the asymptotic decay of $\hat{\delta}^{\ktt_2}_\sigma(\tau)$ dominates the rise in $\hat{\delta}^{\ktt_1}_\varepsilon(\tau)^{-1}$.
For this reason, a similar picture to that of Sec.~\ref{subsec:analytic_kernel_transformations} emerges.
For instance, an input Cauchy-smeared spectral function $\rho_\varepsilon^{\ctt}(\omega)$ can always be mapped to a Gaussian-smeared $\Gamma^{\gtt}_{\sigma}$.
Likewise, an input Gaussian-smeared spectral function $\rho_\sigma^{\gtt}(\omega)$ can be mapped to a Gaussian-smeared $\Gamma^{\gtt}_{\sigma'}$ provided $\sigma' > \sigma$, and similarly for the Cauchy-to-Cauchy case.

Another possible approach to smoothing \cref{eq:integral_finite_domain} comes from directly smearing the step functions.
Introducing a regularized Heaviside function $\vartheta_\sigma$, with common examples including the logistic or error functions
\begin{align}
    \vartheta^{\rm logistic}_\sigma(x) = \frac1{1+e^{-x/\sigma}} \,, \quad
    \vartheta^{\rm erf}_\sigma(x) = \frac{1 +\mathrm{erf }(x/\sigma)}{2} \,,
\end{align}
leads to the smeared window
\begin{align}
    W^{\mathtt x}_\sigma(\omega;\omega_0, \omega_1) =
    \vartheta^{\mathtt x}_\sigma(\omega-\omega_0) \vartheta^{\mathtt x}_\sigma(\omega_1-\omega),
\end{align}
with $\mathtt x \in \{ \mathrm{logistic}, \mathrm{erf} \}$.
It is also common only to smear the step function localized at the kinematic threshold $\omega_1$, relying on integration against $\rho(\omega)$ to remove contributions from $\omega < \omega_0$.
Either case leads to the smoothed version of the integral in \cref{eq:integral_finite_domain} as
\begin{align}
    \Gamma^{\mathtt x}_\sigma =
    \int d\omega \, R(\omega) W_\sigma^{\mathtt x}(\omega;\omega_0, \omega_1) \rho(\omega) \,.
    \label{eq:I_sigma_smeared_window}
\end{align}

The derivation of \cref{eq:linear_map_to_finite_integral} relies crucially on the convolution theorem.
The choice of the smeared window in \cref{eq:I_sigma_smeared_window} complicates a similar explicit discussion, since the product $R(\omega) W_\sigma^{\mathtt x}(\omega;\omega_0, \omega_1)$ amounts to the convolution $(\hat R * \hat W_\sigma^{\mathtt x})(\tau; \omega_0, \omega_1)$ in Fourier space, rather than a simpler product.
Although the calculation now depends on the functional form of the rational function, the general picture remains unchanged, as can be explicitly verified e.g., for $R(\omega)=1$.
The Fourier-space damping implied by the smoothed window function must dominate the growth of $\hat{\delta}^{\ktt_1}_\varepsilon(\tau)^{-1}$ in order to define the corresponding transition kernel.

\paragraph{Regulated principal-value kernels.}
Regulated versions of the principal value (P.V.) prescription constitute another  class of phenomenologically interesting smearing kernels, which arise naturally in nonperturbative applications of the LSZ formula~\cite{Bulava:2019kbi,DiCarlo:2025mnm}.
The associated kernel is translationally invariant and has the form
\begin{equation}\label{eq:PV-kernel}
    \Pi_\varepsilon(\omega-\omega') \equiv \real \left[ \frac1{\omega-\omega'+i\varepsilon} \right]
    = \frac{\omega-\omega'}{(\omega-\omega')^2+\varepsilon^2} \,.
\end{equation}
The standard P.V. prescription is recovered from the $\varepsilon\to 0$ limit of the integral of this kernel with a test function.

For concreteness, consider the transition map from a Cauchy kernel of width $\varepsilon_1$ to a P.V. kernel of width $\varepsilon_2$.
Translational invariance allows \cref{eq:K_cauchy_to_general} to be applied directly.
The Fourier transform of \cref{eq:PV-kernel} is:
\begin{equation}
    \hat\Pi_{\varepsilon_2}(\tau) = -\frac{i}{2} e^{-\varepsilon_2 \vert\tau\vert}
    \left[ \vartheta(\tau) - \vartheta(-\tau) \right] \,.
\end{equation}
Therefore the transition map is given by
\begin{equation}
    \Pi_{\varepsilon_2-\varepsilon_1}(\omega-\omega')
    =
    \frac{\omega'-\omega}{(\omega'-\omega)^2+(\varepsilon_2-\varepsilon_1)^2} \,.
\end{equation}
Thus, a transition to a P.V. kernel of width $\varepsilon_2$ from a Cauchy kernel of width $\varepsilon_1$ is accessible whenever $\varepsilon_2>\varepsilon_1$. 
This result is also easily interpretable in terms of the analytic properties of the input and smearing kernels (see the discussion below \cref{eq:K_cauchy_to_general}), which are, respectively, related to the imaginary and real parts of the same function, cf. \cref{eq:PV-kernel}.
For the same reason, this discussion generalizes straightforwardly to the other kernel transformations considered above.
Given the identical asymptotic decay of the P.V. and Cauchy kernels in Fourier space, the convergence conditions are analogous to those derived for transitions involving the Cauchy kernel.

\subsection{Regulated kernel transformations}\label{subsec:formalism-regulated}
\noindent
The discussion so far has mostly focused on exact kernel transformations that exist unambiguously.
However, the results of Sec.~\ref{subsec:analytic_kernel_transformations} already presented instances of maps where the desired operator does not always exist as a continuum object without intermediate regularization.
This extends to instances of inverse problems from Euclidean correlation functions as input data. 
To render these transitions tractable, one possibility is to replace the target kernel $\kappa$ by an approximation $\kappa_T$ that can actually be realized from the input family.
The observable $\rho[\kappa_T]$ can then be exactly extracted as a smeared observable for the modified kernel.
The difference $\rho[\kappa]-\rho[\kappa_T]$ represents a bias that must be quantified, for instance with bounds.
Moreover, even when an exact operator exists in the continuum, it can be practically convenient to make this replacement for numerical applications.

For example, an approximation can be explicitly realized by regularizing any of the above integrals in Fourier space as
\begin{align}\label{eq:regularization_Fourier}
    \int_{\R} d\tau \mapsto \int_\R d\tau F_T(\tau) \,,
\end{align}
where the cutoff function $F_T(\tau) \in [0,1]$ satisfies $F_T(\tau) \to 1$ for $\vert\tau\vert \ll T$ and $F_T(\tau) \to 0$ for $\vert\tau\vert \gg T$.
This treatment defines regulated transition kernels, generically denoted $K^{\kappa\leftarrow\phi}_T$ to mirror \cref{eq:kernel_transformation}, and thus
\begin{align}\label{eq:kernel_transformation_T}
    \rho[\kappa] = \lim_{T\to\infty} \rho[\kappa_T] \,, \quad
    \rho[\kappa_T] = \int_A d\alpha \, K^{\kappa\leftarrow\phi}_T(\alpha) \, \rho[\phi_\alpha] \,.
\end{align}
The cutoff function $F_T(\tau)$ can be realized in different ways, e.g., through a hard or smooth Heaviside $\vartheta(T-\vert\tau\vert)$ function.

Alternative approximations are possible, for instance by discretizing the energy-space,
\begin{align}
    \int d\omega \, f(\omega) \to \sum_i \Delta \omega_i f(\omega_i) \,,
\end{align}
which leads immediately to the approximations\footnote{The vector $[K_\kappa]_{i}$ also depends on any additional parameters defining the target kernel $\kappa$.
The present focus is solely on the internal index $i$ related to discretized energies.}
\begin{align}
    \rho[\phi_{\alpha_j}] = \int d\omega \, \phi_{\alpha_j}(\omega) \rho(\omega)
    \approx \sum_i [K_\phi]_{ji} \rho(\omega_i) \,, \\
    \rho[\kappa] = \int d\omega \, \kappa(\omega) \rho(\omega)
    \approx \sum_i [K_\kappa]_{i} \rho(\omega_i) \,,
\end{align}
in terms of the finite matrix $[K_\phi]_{ji} = \phi_{\alpha_j}(\omega_i)$ and the finite vector $[K_\kappa]_{i} = \kappa(\omega_i)$.
The regulated transition operator $K^{\kappa\leftarrow\phi}$ is then given by
\begin{align}\label{eq:kernel_transformation_T_discrete}
    \rho[\kappa_T] = K_T^{\kappa\leftarrow\phi} \rho[\phi] \,, \quad
    K_T^{\kappa\leftarrow\phi} \equiv K_\kappa K_\phi^{-1} \,,
\end{align}
where the summation over matrix indices is left implicit.
Reference~\cite{Tsuji:2026zku} proposes a similar strategy for the case of $\phi_{\alpha_j}(\omega_i) = e^{-\alpha_j\omega_i}$, with useful considerations on the singular value decomposition of the associated $K_\phi$.

Regardless of the details of its realization, any regularized transition kernel effectively defines a \emph{different} smearing kernel $\kappa_T$, with which the spectral function is smeared exactly.
For instance, $\kappa_T(\omega)$ is implicitly defined by \cref{eq:kernel_transformation_T}.
By first inserting the cutoff function $F_T(\tau)$ into \cref{eq:transition_kernel_general,eq:K_k1k2}, integrating gives
\begin{align}
    \kappa_T(\omega) = \int_{\R} d\tau F_T(\tau) \hat\kappa(\tau) e^{+i\tau\omega} \,.
\end{align}
Likewise, the approximate kernel in \cref{eq:kernel_transformation_T_discrete} takes the form
\begin{align}
    \kappa_T(\omega) = \sum_{i} [K_\kappa]_i \delta_T(\omega_i, \omega) \,,
\end{align}
where $\delta_T$ defines a regularized Dirac $\delta$ function:\footnote{Notice that $\delta_T(\omega_i,\omega_j) = \delta_{ij}$ when evaluated at the grid point $\omega=\omega_j$, but this is not true for a generic $\omega$ at fixed regulator (energy spacing).
As the regulator is removed, i.e., as the grid is taken to be dense, $\delta_T$ approaches the Dirac $\delta$ function in the usual distributional sense.
}
\begin{align}
    \delta_T(\omega_i, \omega) = \sum_j [K_\phi^{-1}]_{ij} \phi_{\alpha_j}(\omega) \,.
\end{align}
Note that this is a common aspect of many approaches to spectral reconstructions, including the Backus-Gilbert ~\cite{backus_gilbert,Hansen:2017mnd} and Hansen-Lupo-Tantalo (HLT)~\cite{Hansen:2019idp,Lupo:2026vdj} methods, Chebyshev polynomials~\cite{Bailas:2020qmv}, and Mellin-based methods~\cite{Bruno:2024fqc,Giusti:2026mcy}.
These strategies realize different approximations to $\kappa_T(\omega)$ and $\delta_T(\omega_i, \omega)$; see, e.g., Ref.~\cite{Bruno:2024fqc} for explicit examples.
The key question is the extent to which the bias induced by the regularization,
\begin{align}\label{eq:bias_regulated}
    \mathrm{bias} = \rho[\kappa] - \rho[\kappa_T] \,,
\end{align}
can be successfully estimated or reliably bounded.

In anticipation of the discussion in later sections, it bears emphasizing that the overall uncertainty associated with a regulated reconstruction has two conceptually distinct origins. 
First, there is the ``kinematic mismatch" associated with the fact that $\kappa_T(\omega) \neq \kappa(\omega)$ which leads to the bias in \cref{eq:bias_regulated}.
Second, there is the ``dynamical uncertainty" associated with incomplete knowledge of the underlying spectral function.
In cases of practical interest, general approximation arguments, including those that invoke the Stone-Weierstrass theorem, can be used to establish the convergence of $\kappa_T(\omega)\to \kappa(\omega)$, suggesting that the kinematic mismatch $\rho[\kappa-\kappa_T]$ can be made arbitrarily small under suitable assumptions; see, e.g., Appendix E of Ref.~\cite{Patella:2024cto}.
The dynamical uncertainty, instead,  stems from the challenge of inferring spectral information unambiguously from limited data, and is 
related to the epistemic uncertainty inherent in defining what is meant by ``analytic continuation'' from a finite set of points.
Whenever finite regularization is required for definition of a kernel transformation, one expects a finite dynamical uncertainty.
This dynamical uncertainty is present even for exact data and has been fully  quantified analytically~\cite{Abbott:2025snz,Bergamaschi:2023xzx} and numerically~\cite{Lawrence:2024hjm,Abbott:2026wdw,Mutzel:2026vyw} in several different cases.

\subsection{Bias in regulated kernel transformations}\label{subsec:formalism-bias}
\noindent
Regulated transformations reduce the remaining problem to that of quantifying the associated kernel-mismatch bias.
A precise quantification of the bias is often inaccessible in practice, but the input data alone can still suffice to estimate an upper bound.
For any non-negative spectral density $\rho$ and positive smearing kernel $\kappa$, H{\"o}lder's inequality implies
\begin{align}
    \rho[\kappa] - \rho[\kappa_T]
    &= \int_0^\infty d\omega \, \rho(\omega) \left[ \kappa(\omega) - \kappa_T(\omega) \right]
    \\
    \label{eq:bound_rho_kappa_from_data}
    &\leq \left\Vert \frac{ \kappa(\omega) - \kappa_T(\omega) }
    { \phi_\alpha(\omega) } \right\Vert_\infty \rho[\phi_\alpha]
\end{align}
for any arbitrary positive kernel function $\phi_\alpha$, with the inequality holding for any fixed value of $\alpha$.
The norm is to be understood as the pointwise maximum of the absolute value of its argument.\footnote{
More formally, $\Vert f(\omega) \Vert_\infty = \sup_{\omega} | f(\omega) |$.
Since the difference between $\inf/\min$ and $\sup/\max$ plays no important role in the applications at hand, this distinction is omitted here and elsewhere in the main text.
}

Having factorized the dynamical and kinematical information, respectively encoded in the spectral and smearing functions, the right-hand side of \cref{eq:bound_rho_kappa_from_data} can now be quantified directly from the input data for $\rho[\phi_\alpha]$.
It also encodes the expected asymptotic behavior, with the bound on the bias decreasing as the kernel approximation improves. 
Alternative bounds, specific to particular reconstruction methods, exist in certain cases including Ref.~\cite{Kellermann:2025pzt,Bruno:2023bue,Bruno:2024fqc,Tsuji:2026zku}.
In such instances, \cref{eq:bound_rho_kappa_from_data} constitutes a complementary estimate.

The bound in \cref{eq:bound_rho_kappa_from_data} is certainly not necessarily tight.
Given the freedom in choosing $\alpha$, it might be useful to minimize this bound with respect to $\alpha$.
However, this line of thinking can only be of practical use when the norm is not divergent or too large compared to $\rho[\kappa_T]$.
In particular, the ratio $\kappa(\omega) / \phi_\alpha(\omega)$ should not diverge in the first place.
For example, this tail-compatibility requirement excludes Gaussian-to-Cauchy targets but allows Gaussian-to-Gaussian cases that are not feasible through the exact analytic transformations defined in Sec.~\ref{subsec:analytic_kernel_transformations}.

For the case of input Euclidean time correlation functions, where $\phi_t(\omega) = e^{-\omega t}$, \cref{eq:bound_rho_kappa_from_data} furnishes upper bounds on the bias for a target Gaussian smearing kernel, but not for a Cauchy one.
Alternative smearing kernels might be targeted instead, allowing a proper quantification of the bias.
For example, one could use
\begin{align}\label{eq:Cauchy-like}
    \delta_\varepsilon^{\ctt}(\omega,\omega';t_0) =
    t_0 e^{-\omega t_0} \delta_{\varepsilon t_0}^{\ctt}(e^{-\omega t_0}, e^{-\omega' t_0}) \,,
\end{align}
which recovers a standard Cauchy kernel in \cref{eq:delta_cauchy} in the $t_0 \to 0$ limit, while still satisfying $\lim_{\varepsilon\to0} \delta_\varepsilon^{\ctt}(\omega,\omega';t_0) = \delta(\omega-\omega')$, since the prefactors of the modified Cauchy kernel precisely cancel those coming from the composition rule of the Dirac delta function for the argument $\phi_{t_0}(\omega)=e^{-\omega t_0}$.

Although \cref{eq:bound_rho_kappa_from_data} is the main kernel-level bound used later, it is not the only possible estimate.
As a secondary illustration, the Fourier-space cutoff in \cref{eq:regularization_Fourier}  provides an alternative strategy to quantify the bias.
For the sake of concreteness, consider a Gaussian-to-Gaussian transformation.
In Fourier space, the smeared spectral function takes the form
\begin{align}
    \hat{\rho}_\sigma^{\gtt}(\tau) = e^{-\tfrac12 \sigma^2 \tau^2} \hat{\rho}(\tau) \,,
\end{align}
with its regularized version being $F_T(\tau) \hat{\rho}_{\sigma}^{\gtt}(\tau)$.
The energy-space bias in \cref{eq:bias_regulated} is precisely that induced by the cutoff function,
\begin{align}
    \rho_{\sigma}^{\gtt}(\omega) - [{\rho}^{\gtt}_{\sigma}]_T(\omega) =
    \int_{\R} d\tau \, (1 - F_T(\tau)) \, \hat{\rho}^{\gtt}_{\sigma}(\tau) \, e^{i\tau\omega} \,.
\end{align}
This expression immediately implies a pointwise bound on the bias:
\begin{align}
    \norm{\rho_{\sigma}^{\gtt} - [{\rho}_{\sigma}^{\gtt}]_T}_\infty
    &\leq \int_{\R} d\tau \, \abs{1-F_T(\tau)} \, \abs{\hat{\rho}^{\gtt}_{\sigma}(\tau)} \\
    &= \int_{\R} d\tau \, \abs{1-F_T(\tau)} \, e^{-\frac12 \sigma^2 \tau^2} \, \abs{\hat{\rho}(\tau)} \,.
\end{align}
In the case of a hard cutoff $F_T(\tau) = \vartheta(T-|\tau|)$, H{\"o}lder's inequality further implies
\begin{align}
    \norm{\rho_{\sigma}^{\gtt} - [{\rho}_{\sigma}^{\gtt}]_T}_\infty
    \leq
    \norm{\rho}_{1} \frac{1}{\sqrt{2\pi\sigma^2}}
    \mathrm{erfc} \left( \frac{\sigma T}{\sqrt{2}} \right) \,.
\end{align}
Here the bound uses $\rho(\omega) \geq 0$ together with the estimate:
\begin{align}
    \norm{\hat\rho}_\infty
    &= \max_\tau \left\vert \int_0^\infty \frac{d\omega}{2\pi} \rho(\omega) e^{-i\omega\tau} \right\vert \\
    &\leq \frac1{2\pi} \int_0^\infty d\omega \, \rho(\omega) = \frac1{2\pi} \norm{\rho}_1 \,,
\end{align}
which in turn might be quantifiable, for example from sum rules.
In other words, the cutoff smeared spectral function converges exponentially as the cutoff is removed.
Similar results can be derived for other kernel transformations.
Note that this bound is uniform across different centers of the target Gaussian smearing kernel, whereas the bound in \cref{eq:bound_rho_kappa_from_data} is specific to a given center.

%% file: sec_bounds.tex
\noindent
This Section is dedicated to the problem of propagating bounds in kernel transformations, leveraging the positivity of the underlying spectral function.
Assume that pointwise upper and lower bounds are available for the input smeared quantities,
\begin{align}\label{eq:rho_bounds_alpha}
    \rho[\phi_\alpha]_- \leq \rho[\phi_\alpha] \leq \rho[\phi_\alpha]_+ \,.
\end{align}
The goal is to construct rigorous bounds on a target observable $\rho[\kappa]$, either after an exact transformation such as \cref{eq:kernel_transformation} or after a regulated replacement of the target kernel.
If the transition kernels were non-negative, this propagation would be immediate by integrating the lower and upper input bounds against the same positive weights.
However, as \cref{fig:cauchy_to_gaussian_kernel} shows, transition kernels are generically oscillatory.

This Section develops two complementary responses to handle these signed integrands.
The first is a Riesz-Kantorovich (RK) construction, described in Sec.~\ref{subsec:RK_bounds}, that propagates the input intervals directly through the signed transition kernel.
It is inexpensive, general, and rigorous, but can be overly conservative because each input point used to compute such bounds may saturate its interval independently, even for large signed fluctuations.
The second, detailed in Sec.~\ref{subsec:positivity_bounds}, is an optimized construction that explicitly enforces the non-negativity of the underlying physical spectral function.
It can produce improved and potentially tight rigorous bounds, at the cost of a numerical optimization and a separate bias-correction step, described in Sec.~\ref{subsec:positivity_certification} when continuum inequalities are enforced on a grid.
\cref{sec:asymptotics} then discusses how the size of propagated bounds changes and can be optimized with respect to the input smearing. Finally, rigorous certification and optimization of bounds on regulated kernel transformations are included in Sec.~\ref{subsec:RK_bounds_regulated}.

\subsection{Riesz-Kantorovich bounds for kernel transformations}\label{subsec:RK_bounds}
\noindent
For concreteness, this subsection restricts the discussion to kernel transformations defined with respect to energies, as for Cauchy or Gaussian input smearing kernels, with
the same sign-splitting logic applying to other kernel transformations.
Suppose that pointwise non-negative upper and lower bounds are known for a smeared spectral function,
\begin{align}\label{eq:rho_bounds_positive}
    0 \leq
    \rho_{\varepsilon-}^{\ktt}(\omega)
    \leq
    \rho^{\ktt}_{\varepsilon}(\omega)
    \leq
    \rho_{\varepsilon+}^{\ktt}(\omega)
    \,,
\end{align}
for some positive kernel $\ktt$ with parameter $\varepsilon$, e.g., $\ktt \in \{ \ctt, \gtt \}$ as in \cref{eq:delta_cauchy,eq:delta_gaussian}.
The goal is to establish bounds
\begin{align}\label{eq:Krho-bounds-def}
    [K\rho_\varepsilon^\ktt]_- \leq K\rho_\varepsilon^\ktt \leq [K\rho_\varepsilon^\ktt]_+ \,
\end{align}
on the quantity (cf.~\cref{eq:K_k1k2_def})
\begin{equation}\label{eq:Krho-def}
    K\rho_\varepsilon^\ktt = \int_{\mathbb R} d\omega \, K(\omega) \rho^{\ktt}_{\varepsilon}(\omega) \,,
\end{equation}
where $K(\omega)$ denotes a transformation kernel.
Note that $K(\omega)$ might additionally depend on other variables, such as $\omega'$ or $\varepsilon_2$ in \cref{eq:kernel_transformation_general}; such dependence plays no important role in the present discussion and will be left implicit.

The general sign-splitting argument is local in $\omega$ and intuitive to understand. 
The goal is to minimize cancellations conservatively in the integrand.
For the upper bound on the target, the upper bound $\rho^\mathtt{k}_{\varepsilon+}(\omega)$ on the input is used whenever $K(\omega)$ is positive.
Likewise, the lower bound $\rho^\mathtt{k}_{\varepsilon-}(\omega)$ on the input is used whenever $K(\omega)$ is negative.
Reversing these choices gives the lower bound on the target.
Equivalently, the RK formulas are (cf. Ref.~\cite{AliprantisBurkinshaw2006} and references therein),
\begin{align}\label{eq:lower_bound_Krho}
    [K\rho_\varepsilon^\ktt]_- &= \int_{\mathbb R} d\omega \, \min (K \rho_\varepsilon^\ktt)(\omega) \,,
    \\
    \label{eq:upper_bound_Krho}
    [K\rho_\varepsilon^\ktt]_+ &= \int_{\mathbb R} d\omega \, \max (K \rho_\varepsilon^\ktt)(\omega) \,,
\end{align}
where the integrands are defined by the following expressions:
\begin{align}
    \label{eq:min_Krho}
    \min (K \rho_\varepsilon^\ktt)(\omega)
    &\equiv\begin{cases}
        K(\omega) \rho_{\varepsilon-}^\ktt(\omega) \,, & K(\omega) \geq 0 \\
        K(\omega) \rho_{\varepsilon+}^\ktt(\omega) \,, & K(\omega) < 0
    \end{cases}
    \\
    \label{eq:max_Krho}
    \max (K \rho_\varepsilon^\ktt)(\omega)
    &\equiv\begin{cases}
        K(\omega) \rho_{\varepsilon+}^\ktt(\omega) \,, & K(\omega) \geq 0 \\
        K(\omega) \rho_{\varepsilon-}^\ktt(\omega) \,, & K(\omega) < 0 \,.
    \end{cases}
\end{align}
A more thorough, self-contained derivation of these bounds is detailed in Appendix~\ref{app:RK_bounds}.
Notice that, while the bounds in \cref{eq:lower_bound_Krho,eq:upper_bound_Krho} hold under general smoothness requirements on the transition kernels and the underlying smeared spectral integrals, their pointwise realizations in \cref{eq:min_Krho,eq:max_Krho} depend on the positivity assumption on the bounds in \cref{eq:Krho-bounds-def}, with straightforward modifications to more general cases.

\begin{figure*}[t!]
    \centering
    \includegraphics[width=\linewidth]{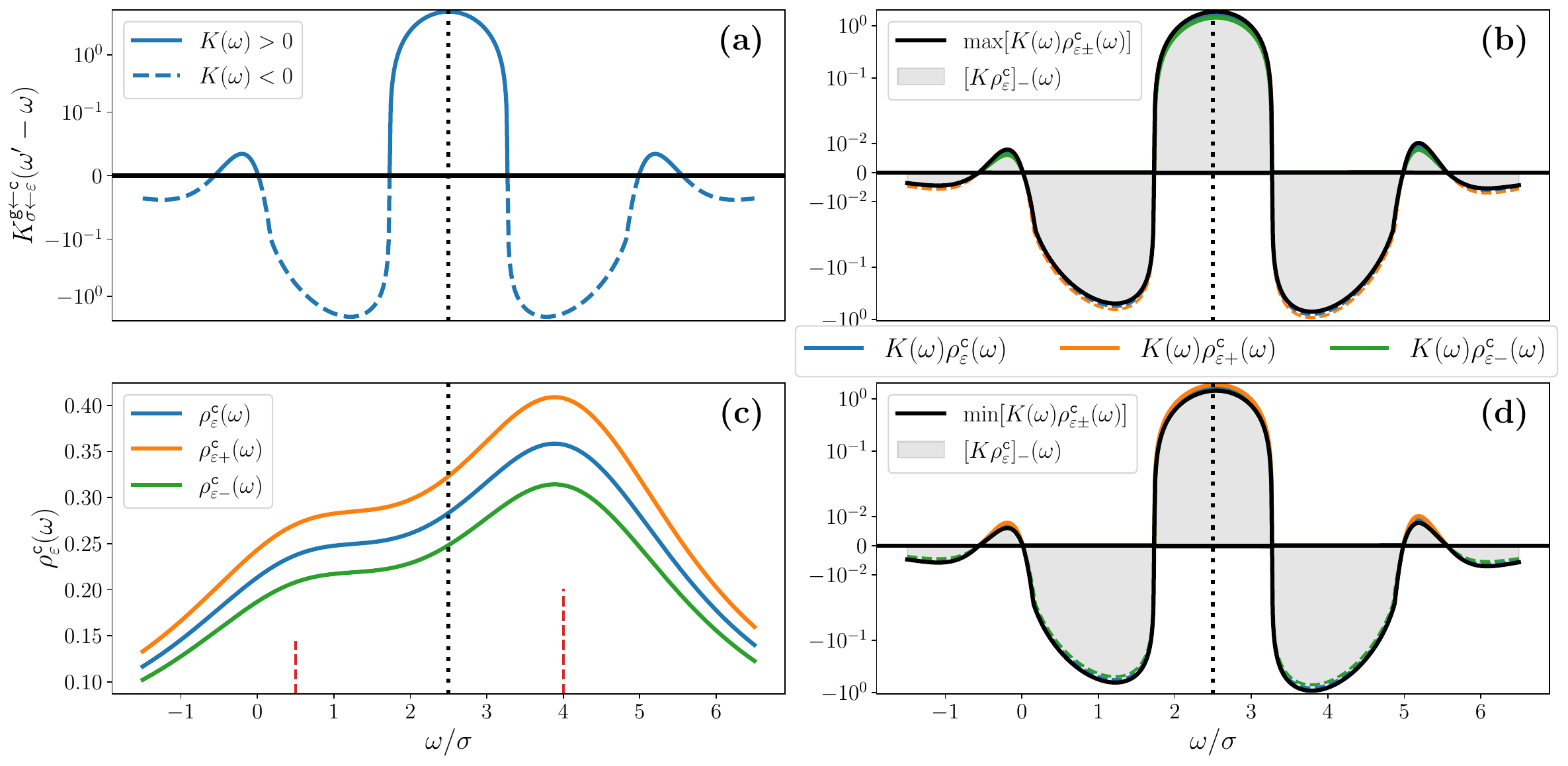}
\caption{RK bounds for the transition kernel $K_{\sigma\leftarrow\varepsilon}^{\gtt \leftarrow \ctt}(\omega-\omega')$, cf. \cref{eq:kernel_transformation_general}.
Panel (a) shows the transition kernel $K_{\sigma\leftarrow\varepsilon}^{\gtt \leftarrow \ctt}(\omega-\omega')$ for fixed $\varepsilon$, $\sigma$, and $\omega'$, with the solid and dotted lines highlighting the regions where the kernel is positive and negative, respectively.
Panel (c) shows a toy spectral function, cf. \cref{eq:toy_spectral_function}, smeared with a Cauchy kernel, alongside artificial upper and lower bounds defined in \cref{eq:toy_bounds}.
The vertical red dotted lines indicate the position of the two peaks in $\rho$, with heights proportional to their amplitudes.
Panels (b) and (d) depict the integrands contributing to the RK upper and lower bounds, respectively, cf. \cref{eq:max_Krho,eq:min_Krho}.}
    \label{fig:RK_bounds}
\end{figure*}

\Cref{fig:RK_bounds} provides a visual construction of the RK bounds for the transition kernel $K_{\sigma\leftarrow\varepsilon}^{\gtt \leftarrow \ctt}(\omega-\omega')$.
The example uses a toy spectral function,
\begin{equation}\label{eq:toy_spectral_function}
    \rho(\omega) = \delta(\omega-\tfrac{1}{2}\sigma) + 2 \delta(\omega-4\sigma)
\end{equation}
smeared with a Cauchy kernel, together with artificial upper and lower bounds of the form
\begin{equation}\label{eq:toy_bounds}
    \rho_{\varepsilon\pm}^{\ctt}(\omega)
    = \rho_{\varepsilon}^{\ctt}(\omega) \left( 1 + \frac{\varepsilon_0}{\varepsilon} \right)^{\pm2},
\end{equation}
where $\varepsilon_0 \geq 0$ is taken to be constant.
The primary role of \cref{eq:toy_bounds} is to define an example that can be used to demonstrate how bounds are propagated, and not to create a realistic model for lattice data.
Although the particular functional form is chosen for convenience and simplicity, the input bounds defined in \cref{eq:toy_bounds} are inspired by rigorously computed bounds on smeared spectral functions extracted from Euclidean data~\cite{Abbott:2025snz}, with which they share several useful qualitative properties:
\begin{enumerate}
\item Positivity: \cref{eq:rho_bounds_positive} is satisfied for all $\omega$ and all positive $\varepsilon$.
\item Divergence as $\varepsilon\to0^+$: $\rho_{\varepsilon+}^{\ctt}(\omega)-\rho_{\varepsilon-}^{\ctt}(\omega)$ diverges polynomially in $1/\varepsilon$, with the lower and upper bounds saturating the trivial bounds of $0$ and $\infty$, respectively.
\item Convergence of the bounds for large $\varepsilon$: as $\varepsilon\to+\infty$, the bounds approach the true result $\rho_{\varepsilon\pm}^{\ctt}(\omega) \to \rho_{\varepsilon}^{\ctt}(\omega)$ polynomially in $\varepsilon$.
\end{enumerate}
Panels (c) and (d) of \cref{fig:RK_bounds} show the integrands that lead to $K\rho_\varepsilon^{\ctt}$ and the RK bounds $[K\rho_\varepsilon^{\ctt}]_\pm$, highlighting the construction of $\max/\min (K\rho_\varepsilon^{\ctt})(\omega)$ within intervals where the transition kernel is positive or negative.
Note that, as expected, the area under the curve for the RK upper bound in panel (b) is strictly larger than the area under the curve for the RK lower bound in panel (d).
In all panels, the vertical black dotted line indicates the fixed position of $\omega'/\sigma$.

For approximate (i.e., regulated) kernel transformations, rigorous bound propagation must be carefully split in two consecutive steps. First, the RK bounds should be computed for the regulated transformation kernel $K_T \approx K$,
\begin{equation}
    [K_T\rho_\varepsilon^\ktt]_- \leq K_T \rho_\varepsilon^\ktt \leq [K_T \rho_\varepsilon^\ktt]_+ \,.
\end{equation}
Subsequently, one must quantify the remaining bias associated with the kernel approximation.
Further details on regulated kernel transformations are provided in Sec.~\ref{subsec:RK_bounds_regulated}.

An important property of kernel transformations can already be appreciated from this simple construction. 
In particular, one could wonder how rigorous pointwise bound propagation compares between a single $\ktt_i \to \ktt_f$ and a chain $\ktt_i \to \ktt_1 \to \dots \to \ktt_{N-1} \to \ktt_f$ of $N>1$ (possibly regulated) kernel transformations. 
One option consists in iteratively propagating pointwise RK bounds on all intermediate $\ktt_n \to \ktt_{n+1}$ kernel transformations. 
While rigorous, this strategy would yield unnecessarily inflated bounds on the final target observable. 
This is due to neglected correlations in intermediate smeared quantities, which necessarily emerge even from pointwise (and hence diagonal) input bounds.
Given the linearity of the transformations, correlations can be maintained to the greatest extent possible
by computing RK bounds directly on the kernel transformation resulting from the composition of all intermediate ones.

\subsection{Positivity-optimized bounds}\label{subsec:positivity_bounds}
\noindent

The RK construction is rigorous and local in the input parameter: each value of $\alpha$ may independently saturate whichever endpoint is most favorable for the signed transition kernel.
This locality is also the source of its conservatism.
If the target and input kernels are positive, positivity of the underlying spectral function guarantees a non-negative target observable and a nontrivial RK upper bound,
\begin{align}
    0
    \leq \int d\alpha \, K(\alpha) \rho[\phi_\alpha]
    \leq [K \rho[\phi_\alpha]]_+ \,,
\end{align}
with $[K\rho[\phi]]_\pm$ defined analogously to \cref{eq:min_Krho,eq:max_Krho} but in terms of the pointwise bounds in $\alpha$, cf. \cref{eq:rho_bounds_alpha}.

However, the same assumptions do not guarantee that the RK lower bound is positive.
Since $\rho[\phi_\alpha]_-$ is only assumed to be non-negative, while $\rho[\phi_\alpha]_+$ can be arbitrarily large, the weighted difference in $\min (K\rho[\phi])(\alpha)$ can in principle achieve negative values.
Depending on the cancellations across different values of $\alpha$, the RK lower bound can either be positive, negative, or zero.
In other words, whether or not the RK lower bound improves on the trivial lower bound ($K \rho[\phi] \geq 0$) depends on the details of the original bounds for $\rho[\phi]$ and on the size of the fluctuations of the transition kernel $K$.

A stronger construction is possible that naturally incorporates the positivity assumptions (when they hold) to tighten the above bounds.
Instead of allowing each input pointwise bound to be saturated independently, the largest and smallest possible values of $\rho[\kappa]$ are identified by scanning among all nonnegative spectral densities compatible with all input bounds in \cref{eq:rho_bounds_alpha}.
This construction naturally preserves the physical constraint that the input observables must arise from a single $\rho(\omega)\geq0$.
The general idea is closely related to that of Refs.~\cite{Abbott:2026wdw,Lawrence:2024hjm,Mutzel:2026vyw}.

It is convenient to re-express the input bounds in terms of a central value $\overline\rho[\phi_\alpha]$ and half-width $\Delta\rho[\phi_\alpha]$,
\begin{align}\label{eq:rho_bounds_alpha_mean_error}
    \overline\rho[\phi_\alpha] \equiv \frac{\rho[\phi_\alpha]_+ + \rho[\phi_\alpha]_-}{2} \,, \\
    \label{eq:rho_bounds_alpha_error}
    \Delta\rho[\phi_\alpha] \equiv \frac{\rho[\phi_\alpha]_+ - \rho[\phi_\alpha]_-}{2} \,.
\end{align}
The exact upper bound can then be written as the following optimization over the convex space of admissible positive-semidefinite spectral densities\footnote{
    Additional constraints, for instance imposing particular asymptotic behavior or a suitable notion of smoothness or continuity, may also be imposed as long as they are convex.
}:
\begin{equation}\label{eq:primal-upper}
\begin{aligned}
    &\rho[\kappa]_+^\star
    = \max_{\rho} \quad \rho[\kappa]
    \\
    & \qquad \text{such that} \,\, \left| \rho[\phi_\alpha] - \overline\rho[\phi_\alpha] \right|
            \le \Delta\rho[\phi_\alpha] \quad \forall \alpha \,, \\
    & \qquad \text{and} \,\, \rho(\omega) \geq 0 \quad \forall \omega \in [0, \infty) \,.
\end{aligned}
\end{equation}
The lower bound is obtained analogously by substituting $\rho[\kappa]_-^\star = \min_{\rho} \rho[\kappa]$ in the first line, or equivalently by negating the objective.
The (primal) semi-infinite program (SIP) in \cref{eq:primal-upper} defines the positivity-optimized bounds.

The dual point of view has a simple kernel interpretation.
Suppose a finite linear combination of input kernels,
\begin{equation}\label{eq:kappa_lambda}
    \kappa_\lambda(\omega) = \sum_i \lambda_i \phi_{\alpha_i}(\omega) \,,
\end{equation}
gives an upper bound for the target kernel, $\kappa_\lambda(\omega)\geq\kappa(\omega)$ for all $\omega\geq0$.
For any nonnegative spectral density, this kernel inequality implies $\rho[\kappa_\lambda] \geq \rho[\kappa]$.
The input intervals then bound $\rho[\kappa_\lambda]$ by the central values and half-widths in \cref{eq:rho_bounds_alpha_mean_error}.
Minimizing this upper estimate over all such $\kappa_\lambda$ gives the dual upper bound,
\begin{equation}\label{eq:dual-upper}
\begin{aligned}
    & \rho[\kappa]_+^\star = \min_{\{\lambda_i\}} \,\,
    \sum_i \lambda_i \overline\rho[\phi_{\alpha_i}] + \sum_i |\lambda_i| \cdot \Delta\rho[\phi_{\alpha_i}] \\
    & \qquad \text{such that} \,\, \kappa_\lambda(\omega) \ge \kappa(\omega)
    \quad \forall \omega \in [0, \infty) \,.
\end{aligned}
\end{equation}
The lower bound $\rho[\kappa]_-^\star$ is formulated identically by using kernel lower bounds, taking the maximum over $\{\lambda_i\}$, and reversing the signs of the penalty and continuum inequality.
The derivation of the stated primal-dual equivalence is given in Appendix~\ref{app:sip_dual}.
Note that choice of a particular finite set of input kernels $\{\phi_{\alpha_i}\}$
does not introduce any approximation provided \emph{some} upper bound $\kappa_\lambda(\omega)$ can be identified, whether optimal or not.
That is, the associated bound will generically be conservative.

Similar care as for RK bound propagation must be applied when composing kernel transformations through iterative SIP steps. In particular, while iterative propagation of pointwise intermediate certified intervals preserves rigor, it is generally more conservative than a direct SIP solve for the composed transformation.

The primal and dual problems in \cref{eq:primal-upper,eq:dual-upper} have several important differences with respect to similar problems considered in Refs.~\cite{Lawrence:2024hjm,Abbott:2026wdw,Mutzel:2026vyw}.
First, \cref{eq:primal-upper,eq:dual-upper} are phrased in terms of essentially arbitrary positive input kernels $\phi_\alpha$, extending previous treatments focused on Euclidean correlator data.
Second, the pointwise bounding constraint on the input data in the primal problem \cref{eq:primal-upper} leads to an absolute value in the second term of the dual objective function in \cref{eq:dual-upper}. 
As discussed above, due care must taken when composing several kernel transformations.
This is complementary to previous focuses on statistical uncertainty propagation~\cite{Lawrence:2024hjm,Abbott:2026wdw,Mutzel:2026vyw}.
Such statistical variants can be included in the definition of the the primal problem, cf. \cref{eq:primal-upper-noisy}, but they are not the main focus here.
Finally, as described in the following subsection, certified bounds on approximate solutions to \cref{eq:dual-upper} can be obtained from a variant of \cref{eq:bound_rho_kappa_from_data}, cf. \cref{eq:rho_kappa_pm_rig}, complementing rigorous tight results obtained for certain classes of target smearing kernels~\cite{Abbott:2026wdw}.

\subsection{Numerical certification of positivity-optimized bounds}\label{subsec:positivity_certification}
\noindent
In practice, the kernel inequalities in \cref{eq:dual-upper} are enforced on a finite energy grid.
The resulting solution gives a \textit{candidate} upper bound $\rho[\kappa]_+^{\mathrm{cand}}$ and a candidate bounding kernel $\kappa_{\lambda^\ast_+}(\omega) = \sum_i (\lambda_+^\ast)_i \phi_{\alpha_i}(\omega)$.
Because the grid in $\omega$ is finite, $\kappa_{\lambda_+^\ast}(\omega)$ can violate the continuum inequality $\kappa_{\lambda_+^\ast}(\omega) \ge \kappa(\omega)$ between grid points.
Similarly, the candidate lower bound $\rho[\kappa]_-^{\mathrm{cand}}$ comes from a candidate minorant $\kappa_{\lambda^\ast_-}(\omega)$ that may violate $\kappa_{\lambda_-^\ast}(\omega) \le \kappa(\omega)$ between grid points.

The positivity of the spectral density allows these residual continuum violations to be absorbed \textit{a posteriori} by following a similar strategy to that used for deriving \cref{eq:bound_rho_kappa_from_data}.
Define the normalized maximal violations
\begin{align}\label{eq:certification_SIP}
    \delta_\pm(\alpha) \equiv \max \left[ 0, \, \max_\omega \left(
        \pm \frac{\kappa(\omega) - \kappa_{\lambda_\pm^\ast}(\omega)}{\phi_{\alpha}(\omega)}
    \right) \right]
\end{align}
at a chosen fixed value of $\alpha$.
The upper $(+)$ and lower $(-)$ values may differ due to the signed argument of the maximum; taking both equal to the larger of $\delta_+(\alpha)$ and $\delta_-(\alpha)$ recovers the more conservative, symmetric absolute-value structure of \cref{eq:bound_rho_kappa_from_data}.
In other words, the normalized violations in \cref{eq:certification_SIP} allow corrected kernel bounds to be computed, for all $\omega$, by
\begin{align}
    &\kappa_{\lambda_-^\ast}(\omega) - \delta_-(\alpha)\phi_\alpha(\omega) 
    \leq
    \kappa(\omega)\\
    &\kappa(\omega)
    \leq
    \kappa_{\lambda_+^\ast}(\omega)
    +
    \delta_+(\alpha)\phi_\alpha(\omega) \,.
\end{align}
Bounds on $\rho[\kappa]$ immediately follow by correcting the candidate bounds according to
\begin{align}\label{eq:rho_kappa_pm_rig}
    \rho[\kappa]_\pm^{\mathrm{rig}} = \rho[\kappa]_\pm^{\mathrm{cand}}
    \pm \delta_\pm(\alpha) \rho[\phi_\alpha]_+ \,.
\end{align}
Note that, since only bounds on $\rho[\phi_\alpha]$ are presumed to be known, the second term must include $\rho[\phi_\alpha]_+$ to construct rigorous bounds on both $\rho[\kappa]^{\rm rig}_\pm$.
An analogous situation occurs for the right-hand side of \cref{eq:bound_rho_kappa_from_data}.

As in \cref{eq:bound_rho_kappa_from_data}, the choice of $\alpha$ can be optimized to reduce the size of the overall bounds.
If the kernel inequality in \cref{eq:dual-upper} is verified for all $\omega$, then $\delta_\pm(\alpha)=0$ and the candidate bounds are already rigorous.
This separates the numerical search for good candidate kernels from the mathematical certification of the final interval.
When the certification in \cref{eq:certification_SIP} is handled numerically, it may be a useful visual check to plot the bounding kernels (cf. \cref{fig:c2c-comparison} below).\footnote{
    It is clear that the bounds must be respected only on the support of the spectral function. For the cases where the spectral function is an atomic measure, this condition may make the actual bounding envelope of smearing kernels look visually much different from a more general smooth (e.g., $L^2$) approximation. Combining the two approaches may be of practical use. 
}
The same certification idea may also be useful for related dual approaches such as those of Refs.~\cite{Abbott:2026wdw,Lawrence:2024hjm,Mutzel:2026vyw}.

\subsection{Optimal input smearing and RK error scaling\label{sec:asymptotics}}
\noindent
The preceding constructions leave open a design question: which input smearing should be used to obtain the tightest target interval?
For exact transformations, the RK bounds can be computed at each fixed input smearing width $\varepsilon$, but the resulting interval reflects a competition between the quality of the input bounds and the stability of the transition kernel.
The bounds from \cref{eq:min_Krho,eq:max_Krho} immediately imply that the total width of the RK interval is
\begin{align}
    [K\rho_\varepsilon^\ktt]_+ - [K\rho_\varepsilon^\ktt]_-
    &= \int d\omega \, \abs{K(\omega)} \left(
        \rho_{\varepsilon+}^\ktt(\omega)
        -\rho_{\varepsilon-}^\ktt(\omega)
    \right) \nonumber \\
    \label{eq:RK_aysmptotics}
    &= \norm{K \cdot (\rho_{\varepsilon+}^\ktt
        -\rho_{\varepsilon-}^\ktt)}_1\\
    \label{eq:RK_aysmptotics_2}
    &\leq \norm{K}_1 \cdot \norm{
        \rho_{\varepsilon+}^\ktt
        -\rho_{\varepsilon-}^\ktt
    }_\infty \,,
\end{align}
where the third line follows from H{\"o}lder's inequality, and may be replaced by other versions of the same inequality.
For instance, if both the kernel and the difference of the bounds are square integrable, $L^2$ norms may be used in \cref{eq:RK_aysmptotics_2}.

\Cref{eq:RK_aysmptotics} gives a simple way to understand the limiting behavior of the RK width as $\varepsilon$ is varied.
For illustration, consider a Cauchy-to-Gaussian kernel transformation at fixed Gaussian width $\sigma$.
For large $\varepsilon$, the input bounds $\rho_{\varepsilon\pm}^\ctt(\omega)$ may become increasingly tight, since in this limit $\rho_\varepsilon^\ctt(\omega)$ is primarily sensitive to the overall normalization of the underlying distribution $\rho(\omega)$.
At the same time, the transition kernel becomes increasingly oscillatory, and its $L^1$ norm grows exponentially with $\varepsilon/\sigma$.
The RK width therefore grows at large $\varepsilon$.
For small $\varepsilon$, the transition kernel $K$ reduces to a Gaussian (cf. \cref{fig:cauchy_to_gaussian_kernel}), so that $\norm{K}_1 \to 1$.
The original bounds $\rho_{\varepsilon\pm}^\ctt(\omega)$ are expected to grow in this regime, since the limit $\lim_{\varepsilon \to 0}\rho_\varepsilon^\ctt(\omega) = \rho(\omega)$ approaches the unsmeared distribution.
Together, these two regimes indicate the presence of a minimizing input width for a symmetric interval,
\begin{align}\label{eq:optimal_epsilon}
    \varepsilon^\star = \argmin_\varepsilon
    \left( [K \rho_\varepsilon^\ctt]_+ - [K \rho_\varepsilon^\ctt]_- \right) \,.
\end{align}
Intuitively, $\varepsilon^\star$ is the input smearing that best matches the resolution of a given target kernel under the available input bounds.
The same input-smearing question can also be posed for the positivity-optimized bounds, although the resulting interval widths need not follow the symmetric RK estimate in \cref{eq:RK_aysmptotics}.
\Cref{sec:numerics} discusses this optimization for a Cauchy-to-Gaussian kernel transformation (see \cref{fig:stability_analysis}).
The numerical value of $\varepsilon^\star$ will generally depend on both the target smearing width $\sigma$ and the energy $\omega'$ at which it is evaluated.

\subsection{Optimal bounds for regulated kernel transformations}\label{subsec:RK_bounds_regulated}
\noindent
Regulated transformations use the same bound-propagation tools as exact ones, but require one additional step. 
The target kernel $\kappa$ is replaced by an approximant $\kappa_T$, where $T$ denotes the regulator.
One then propagates the input bounds to the regulated observable $\rho[\kappa_T]$, certifies the mismatch between $\kappa$ and $\kappa_T$, and enlarges the propagated interval by that certificate.

Let
\begin{equation}
    \rho[\kappa_T]_- \leq \rho[\kappa_T] \leq \rho[\kappa_T]_+
\end{equation}
denote the propagated RK or positivity-optimized bounds for the regulated kernel, at fixed choice of regulator.
Define the propagated center and half-width by
\begin{align}
    \overline{\rho[\kappa_T]}
    &\equiv
    \frac{\rho[\kappa_T]_+ + \rho[\kappa_T]_-}{2} \,, \\
    \label{eq:delta_rho_kappa_num}
    \Delta_{\rm prop}(T)
    &\equiv
    \frac{\rho[\kappa_T]_+ - \rho[\kappa_T]_-}{2} \,.
\end{align}
For RK bounds obtained from a transition kernel $K_T$, this half-width is equivalent to propagation of the input half-width defined in \cref{eq:rho_bounds_alpha_error} through the absolute-value kernel, i.e.,
\begin{equation}\label{eq:Delta_prop_T}
    \Delta_{\rm prop}(T)
    =
    \int d\alpha \, |K_T(\alpha)|\, \Delta\rho[\phi_\alpha] \,.
\end{equation}
As the regulator is removed, the large oscillations in $K_T$ tend to cancel in the central value but not in $\Delta_{\rm prop}(T)$.

The mismatch between $\kappa$ and $\kappa_T$ is quantified using the rigorous bound on the kernel-level estimate in \cref{eq:bound_rho_kappa_from_data}.
For any positive input kernel $\phi_\alpha$, define
\begin{equation}
    \Delta_{\rm sys}(T;\alpha)
    \equiv
    \left\Vert
    \frac{|\kappa(\omega)-\kappa_T(\omega)|}{|\phi_\alpha(\omega)|}
    \right\Vert_\infty
    \rho[\phi_\alpha]_+ \,.
\end{equation}
The systematic error used in the final bound can be optimized according to
\begin{equation}
    \Delta_{\rm sys}(T) \equiv \min_\alpha \Delta_{\rm sys}(T;\alpha) \,,
\end{equation}
or any computable upper bound from varying $\alpha$.
The total certified half-width is then
\begin{equation}\label{eq:total_systematic_error}
    \Delta_{\rm tot}(T)
    \equiv
    \Delta_{\rm prop}(T) + \Delta_{\rm sys}(T) \,,
\end{equation}
which gives the rigorous interval
\begin{equation}\label{eq:bounds_regulator_T}
    \overline{\rho[\kappa_T]} - \Delta_{\rm tot}(T)
    \leq
    \rho[\kappa]
    \leq
    \overline{\rho[\kappa_T]} + \Delta_{\rm tot}(T) \,.
\end{equation}
The useful regulator is therefore chosen by minimizing this certified total width,
\begin{equation}\label{eq:optimized_regulator_T}
    T^\star \in \argmin_T \Delta_{\rm tot}(T) \,.
\end{equation}

An alternative strategy, closer to the positivity-optimized construction of \cref{eq:rho_kappa_pm_rig}, consists in tuning the regulator to separately identify the sharpest upper and lower envelopes,
\begin{equation}\label{eq:rho_kappa_pm_rig_RK}
\begin{aligned}
    \max_T & \left\{
        \rho[\kappa_T]_- - \Delta_{\rm sys}(T)
    \right\} \\
    & \leq \rho[\kappa] \leq 
    \min_T \left\{
        \rho[\kappa_T]_+ + \Delta_{\rm sys}(T)
    \right\} \,.
\end{aligned}
\end{equation}
Since the optimal upper and lower certificates may occur at different regulator values, this one-sided optimization is not equivalent to minimizing the total width in \cref{eq:optimized_regulator_T}.
We show numerical results using these two optimization methods in Sec.~\ref{sec:numerics} for a sharpening Cauchy-to-Cauchy regulated kernel transformation (see \cref{fig:c2c-combined}).

Although it is not the focus of this work, the reformulation of RK bound propagation extends naturally to linear propagation of statistical errors.
In particular, assuming the input data have an associated covariance matrix $\Sigma_{\alpha,\alpha'}$, the propagated variance on $\overline{\rho[\kappa_T]}$ is
\begin{equation}
    \Delta_{\text{stat}}^2(T) = \int d\alpha \, d\alpha' \, K_T(\alpha) \Sigma_{\alpha,\alpha'} K_T(\alpha') \,.
\end{equation}
In such cases, one can envision an optimization of the regulator based on a minimization of the total error, including the statistical uncertainty.

The Cauchy-to-Cauchy sharpening example in Sec.~\ref{sec:numerics} uses a Tikhonov realization of this general construction.
It is useful to record here the practical details of the Tikhonov regularization as an explicit instance of regulated kernel transformations with a tunable parameter.
Let $\{\omega_j\}$ be a dense energy grid and let $\{\alpha_i\}$ be the grid of input kernel centers.
For an input Cauchy width $\varepsilon_1$, define the input-kernel matrix and target kernel vector, respectively, as
\begin{align}
    [K_{\varepsilon_1}^{\ctt}]_{j i}
    &\equiv
    \delta_{\varepsilon_1}^{\ctt}(\alpha_i-\omega_j) \,, \\
    [K_{\varepsilon_2}^{\ctt}(\omega')]_j
    &\equiv
    \delta_{\varepsilon_2}^{\ctt}(\omega'-\omega_j) \,.
\end{align}
The coefficient vector $g_{\regulator}(\omega')$ is defined as the solution to the Tikhonov problem
\begin{equation}\label{eq:g_reg1}
    g_{\regulator}(\omega')
    =
    \argmin_g
    \left\{
    \norm{K_{\varepsilon_1}^{\ctt} g-K_{\varepsilon_2}^{\ctt}(\omega')}_2^2
    + \regulator \norm{g}_2^2
    \right\},
\end{equation}
with closed-form solution
\begin{equation}\label{eq:g_reg2}
    g_{\regulator}(\omega')
    =
    \left[ (K_{\varepsilon_1}^{\ctt})^T K_{\varepsilon_1}^{\ctt}+\regulator \mathbb{I} \right]^{-1}
    (K_{\varepsilon_1}^{\ctt})^T K_{\varepsilon_2}^{\ctt}(\omega') \,.
\end{equation}
This defines the regulated target vector
\begin{equation}
    \overline K_{\varepsilon_2,\regulator}^{\ctt}(\omega')
    =
    K_{\varepsilon_1}^{\ctt} g_{\regulator}(\omega') \,,
\end{equation}
or equivalently the regulated kernel $\kappa_{\regulator}$ obtained as a linear combination of input kernels centered at $\alpha_i$.
The corresponding regulated observable is the same linear combination of the input smeared data, and the final bound is obtained from \cref{eq:total_systematic_error}.
Although regularization schemes besides Tikhonov regularization are often used in practice, the details of the regularization are secondary for the present discussion, as long as the associated bias is properly quantified. In this case, a practical realization of a regularized scheme is presented in detail to serve as a concrete example for the construction of a regulated kernel approximation  through a deterministic linear operator controlled by the stabilizing parameter $\regulator$. The latter can then be directly optimized as in \cref{eq:optimized_regulator_T}, with a detailed discussion postponed to Sec.~\ref{subsec:numerics-c2c}.

We conclude this Section with a comment on the \emph{tightness} of the proposed bounds of a target smeared spectral function, subject to two constraints: causality of the underlying spectral density, and the computability of such bounds from the input data $\rho[\phi_\alpha]$.
For simplicity, we first focus on the exact problem, $\Delta\rho[\phi_\alpha]=0$, with a direct generalization including statistical and systematic errors.
Given a linear approximation $\kappa_\lambda(\omega) = \sum_\alpha \phi_\alpha(\omega) \lambda_\alpha$ of the target smearing kernel $\kappa(\omega)$ on the input basis $\phi_\alpha(\omega)$ with coefficients $\lambda_\alpha$, the bound \cref{eq:bound_rho_kappa_from_data} can be improved by identifying the optimal (maximally tight) upper and lower bounds, namely
\begin{gather}\label{eq:Uk}
    U_\kappa \equiv \min_{\eta_\alpha} \left\{
        \sum_\alpha \eta_\alpha \rho[\phi_\alpha] : \sum_\alpha \eta_\alpha \phi_\alpha(\omega) \geq \kappa(\omega) - \kappa_\lambda(\omega)
    \right\} \,, \\
    \label{eq:Lk}
    L_\kappa \equiv \max_{\eta_\alpha} \left\{
        \sum_\alpha \eta_\alpha \rho[\phi_\alpha] : \sum_\alpha \eta_\alpha \phi_\alpha(\omega) \leq \kappa(\omega) - \kappa_\lambda(\omega)
    \right\} \,.
\end{gather}
\Cref{eq:bound_rho_kappa_from_data} amounts to a specific choice of coefficients $\eta_\alpha$: $1$ at a given value of $\alpha$ and $0$ otherwise.
Since the bounds in \cref{eq:Uk,eq:Lk} are the most general possible given the stated hypotheses, they are tight by construction and will typically be tighter than those in \cref{eq:bound_rho_kappa_from_data}.
In fact, linearity allows one to relabel the optimization variables  $\eta_\alpha \mapsto \eta'_\alpha = \eta_\alpha+\lambda_\alpha$. 
This change of variables reveals that the bounds emerging from \cref{eq:Uk,eq:Lk} are the tightest bounds on a smeared positive spectral observable that can be obtained by bounding the target kernel through a linear combination of input basis functions, i.e. by bounding the target smeared spectral function through a linear combination of input data. 
This optimization problem corresponds precisely to the exact dual problem introduced in Ref.~\cite{Abbott:2026wdw}; see also \cref{eq:dual-upper-noisy}.

%% file: sec_numerics.tex
\noindent
This Section provides numerical demonstrations of the framework developed above.
The first three examples use the toy positive spectral function and artificial input bounds of \cref{fig:RK_bounds,eq:toy_spectral_function,eq:toy_bounds} for exact transformations, input-width optimization, positivity-optimized bounds, and certified quantification of the bias in regulated problems.
The first example considers a Cauchy-to-Gaussian transformation, for which the transition kernel is analytically known, cf. \cref{eq:Cauchy_to_Gaussian_exact}.
The second investigates a Cauchy-to-Cauchy sharpening transformation, which requires regularization.
The third example demonstrates the Gaussian-to-Cauchy transformation based on the exact L\'evy mixture in \cref{eq:Levy}.
The final example considers spectral reconstructions as kernel transformations from a noisy Euclidean-time correlation function using regulated exponential-to-Cauchy and exponential-to-Gaussian maps.

\subsection{Analytical kernel transformations (Cauchy to Gaussian)}\label{subsec:numerics-c2g}
\noindent
This first example demonstrates the pipeline for exact transformations.
Starting from the Cauchy-smeared toy data shown in \cref{fig:RK_bounds}, the target is the spectral density $\rho_{\sigma}^\gtt(\omega')$ smeared with a Gaussian kernel at a fixed smearing width $\sigma$, across different energies of the center $\omega'$ of the kernel.
Because the Cauchy-to-Gaussian transition kernel is exact, the bounds require no additional bias quantification.
The example assumes access to bounds for $\rho_{\varepsilon}^\ctt(\omega)$ computed as a function of energy $\omega$ and for a variety of smearing widths $\varepsilon$.

\begin{figure}[t!]
    \centering
    \includegraphics[width=0.99\linewidth]{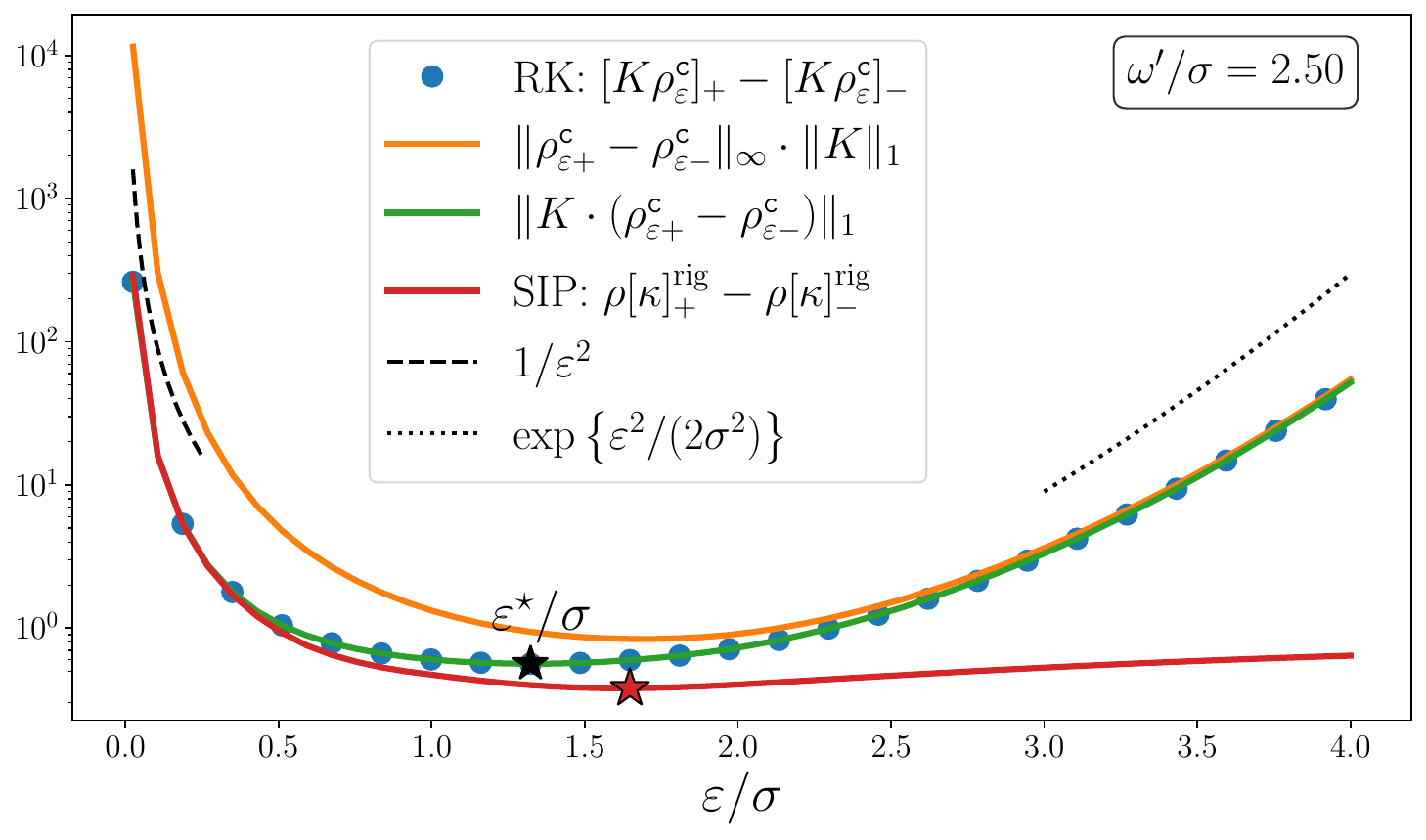}
\caption{Determination of the optimal smearing parameter $\varepsilon^\star$ in \cref{eq:optimal_epsilon} in a Cauchy-to-Gaussian kernel transformation, both for RK and SIP bounds, at a fixed energy $\omega'$ and smearing width $\sigma$ of the target Gaussian kernel.
The details of the model spectral density and the bounds are the same as in \cref{fig:RK_bounds}.}
    \label{fig:stability_analysis}
\end{figure}

The first question is how to choose the input Cauchy width.
The analysis first focuses on a single energy $\omega'$, and aims to determine the optimal input smearing width $\varepsilon^\star$, in the sense of \cref{eq:optimal_epsilon}, that minimizes the width of the estimated bounds.
The bounds are computed both by following the RK prescription, \cref{eq:upper_bound_Krho,eq:lower_bound_Krho}, and by solving the SIP, \cref{eq:dual-upper}, alongside its bias quantification in \cref{eq:certification_SIP}.
The results are shown in \cref{fig:stability_analysis} as a function of $\varepsilon$ for fixed $\sigma$ and fixed $\omega'/\sigma=5/2$.
The behavior of RK bounds for large and small $\varepsilon$ is consistent with the general arguments in Sec.~\ref{sec:asymptotics}.
Unsurprisingly, the optimal value of $\varepsilon^\star$ occurs at an intermediate value $\varepsilon^\star/\sigma = \order{1}$.
The SIP results behave similarly to RK bounds at small $\varepsilon$, due to the divergence of the input bounds, while they increase more mildly at large $\varepsilon$.
For this model, RK bounds are more conservative, due to their generality and the absence of any positivity constraint.
The positivity-optimized optimum lies close to the RK optimum, with a slightly tighter bound.

\begin{figure}[t!]
    \centering
    \includegraphics[width=0.99\linewidth]{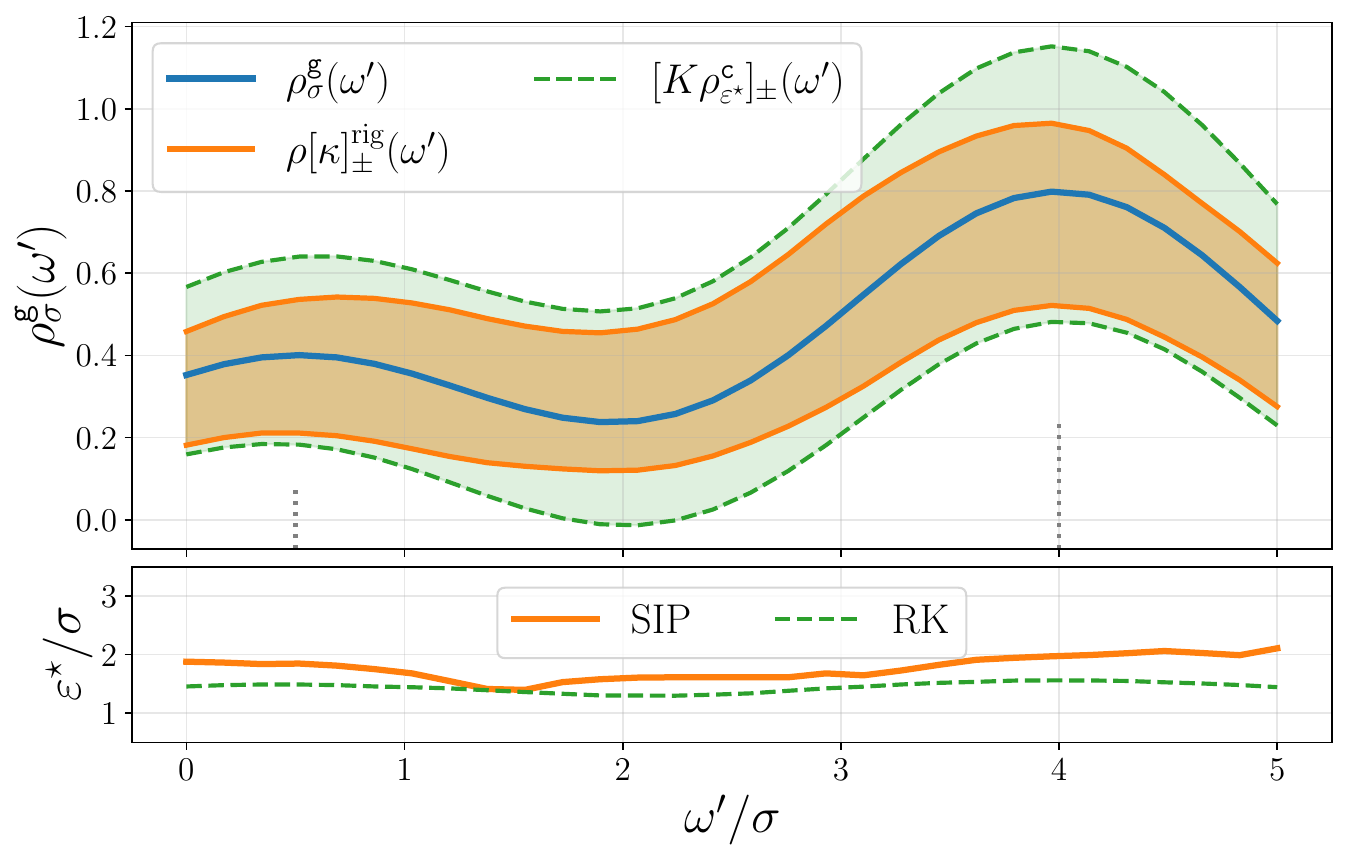}
\caption{Bounds on a Gaussian-smeared spectral function at different values of $\omega'$ at fixed $\sigma$, starting from input Cauchy-smeared spectral functions, see \cref{fig:RK_bounds}.
The vertical dashed lines are the same as in panel (b) of \cref{fig:RK_bounds}, indicating the location and relative sizes of the delta-function peaks in \cref{eq:toy_spectral_function}.
The optimal smearing widths of the input Cauchy kernel $\varepsilon=\varepsilon^\star(\omega')$ are tuned in order to minimize the total uncertainty, as shown in \cref{fig:stability_analysis} for fixed $\omega'$.
Optimal values $\varepsilon^\star(\omega')$ are indicated in the bottom panel for both methods.}
    \label{fig:reconstructions}
\end{figure}

\Cref{fig:reconstructions} shows the result of repeating the analysis in \cref{fig:stability_analysis} for different energies $\omega'$ and fixed target smearing $\sigma$.
The parameter $\varepsilon=\varepsilon^\star(\omega')$ is tuned separately for each value of $\omega'$.
For this toy model, the positivity-optimized intervals remain positive and are tighter than the RK intervals throughout the plotted range.
As discussed above, RK bounds need not be positive because they propagate the pointwise input bands without imposing the existence of a single nonnegative spectral density.
In \cref{fig:reconstructions}, this behavior for the lower bound is present only in a narrow region around $\omega'/\sigma\approx 2$.

A few practical remarks are in order.
First, each reconstruction is performed using the same number of input spectral densities for both strategies for propagating bounds.
The two methods place somewhat different demands on this common input set.
The SIP bounds are most effective when the spectral weight of the smeared input overlaps well with that of the target kernel, whereas the RK bounds require enough coverage to integrate the transition kernel over the full energy range.
To reduce numerical boundary effects, the extraction therefore uses a narrower range of target energies than the full range of input centers for which $\rho_\varepsilon^\ctt(\omega)$ is available.
The main lesson is that the exact Cauchy-to-Gaussian transformation introduces no kernel-mismatch bias; the final bounds are simply the result of propagating the input bounds for the chosen target smearing, subject optionally to a positivity constraint.

\subsection{Regulated kernel transformations (Cauchy to Cauchy)}\label{subsec:numerics-c2c}
\begin{figure}[tpb]
    \centering
    \subfloat[Optimization of the Tikhonov regulator $\regulator$ at fixed target energy $\omega'$.]{
        \includegraphics[width=0.95\columnwidth]{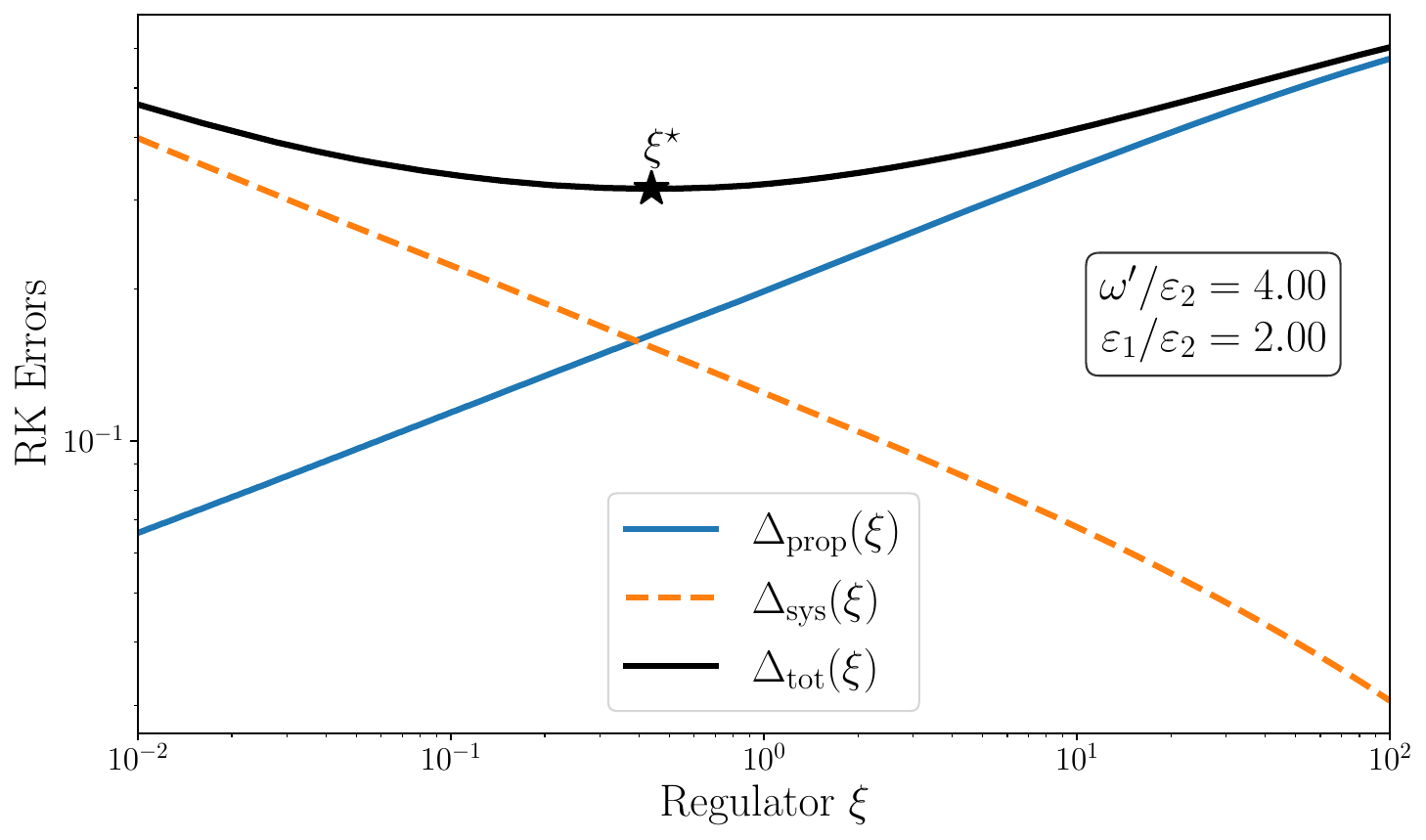}
        \label{fig:c2c-optimization}
    } \\
    \subfloat[Regulated target kernel for sharpened Cauchy reconstructions using RK and SIP methods at fixed target energy $\omega'$.]{
        \includegraphics[width=0.95\columnwidth]{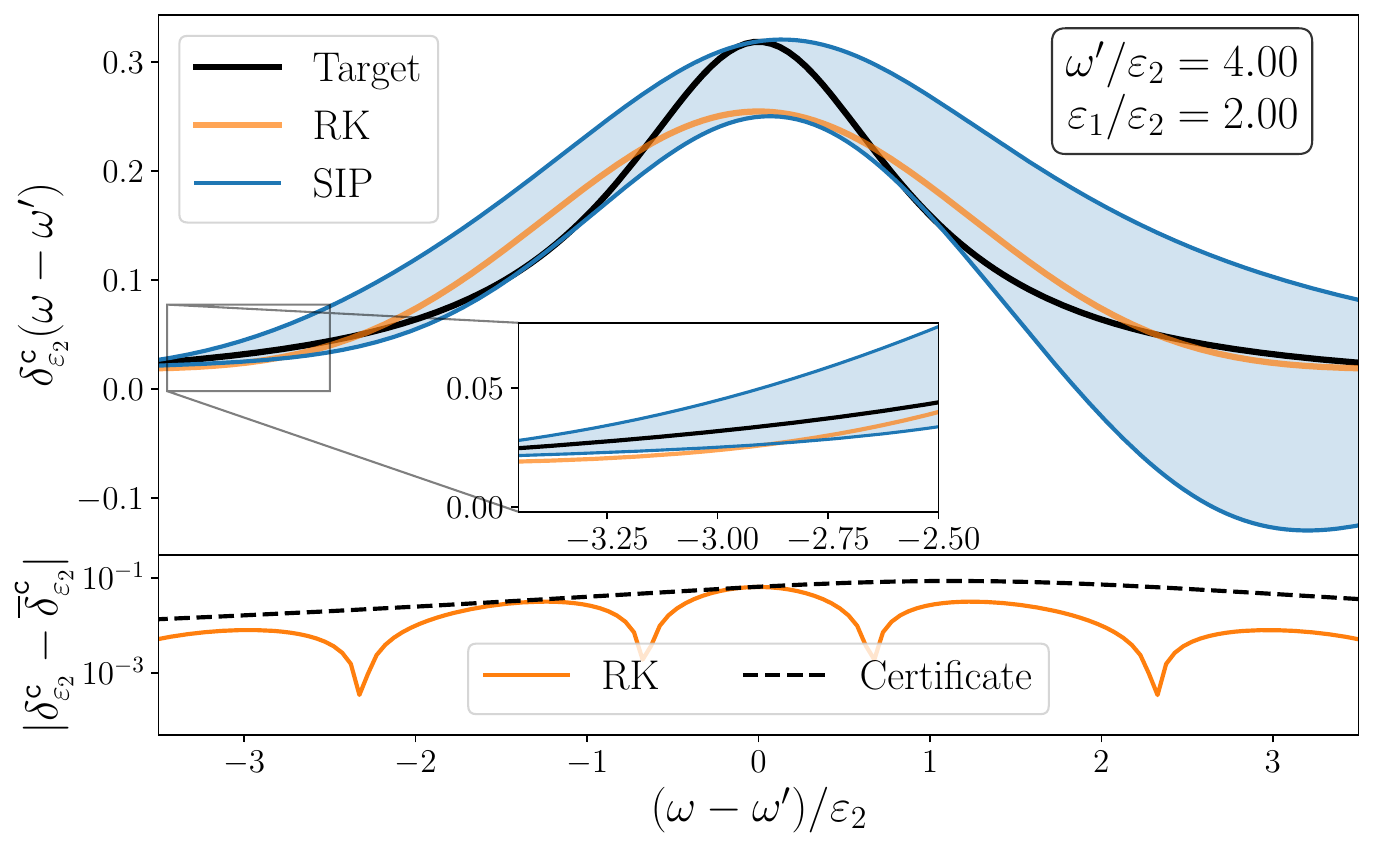}
        \label{fig:c2c-comparison}
    } \\
    \subfloat[Optimized reconstructions at fixed $\varepsilon_1 = 2 \varepsilon_2$ as a function of the target energy $\omega'$.
    The final RK and SIP bounds are shown in green and orange, respectively.
    ]{
        \includegraphics[width=0.95\columnwidth]{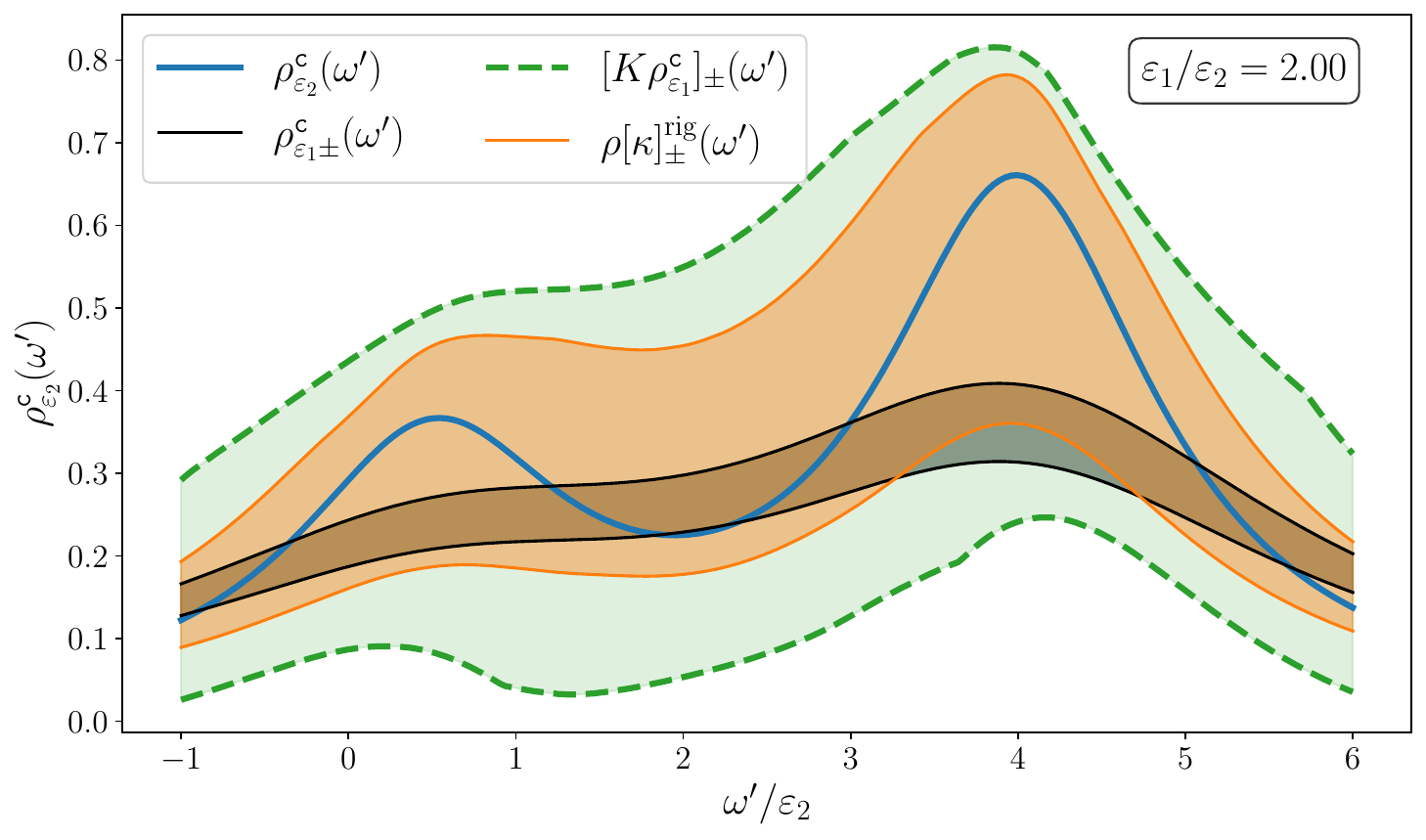}
        \label{fig:c2c-reconstructions}
    }
    \caption{Rigorous Cauchy-to-Cauchy sharpening ($\varepsilon_1 \to \varepsilon_2 = \varepsilon_1/2$) at fixed $\omega'$.
    \Cref{fig:c2c-optimization} shows the choice of the Tikhonov regulator $\regulator$, balancing the propagated half-width $\Delta_{\rm prop}(\regulator)$
    of the RK bounds (orange), cf. \cref{eq:delta_rho_kappa_num}, and the certified kernel-mismatch half-width
    $\Delta_{\rm sys}(\regulator)$ (blue), cf. \cref{eq:bound_rho_kappa_from_data}.
    The total certified half-width $\Delta_{\rm tot}(\regulator)$ of \cref{eq:total_systematic_error} is shown in black, with a minimizing regulator $\regulator^\star$ indicated.
    \Cref{fig:c2c-comparison} illustrates in its top subplot the optimal RK kernel reconstruction, as defined in (a), and the
    corresponding SIP upper/lower bounds. The bottom subplot visualizes the systematic error certificate for the
    target kernel approximation induced at the optimal $\regulator^\star$, together with the data-driven certificate used to
    estimate the associated systematic error.
    Finally, \cref{fig:c2c-reconstructions} demonstrates the reconstructions of the target smeared spectral function at different energies, separately optimizing upper and lower RK bounds, cf. \cref{eq:rho_kappa_pm_rig_RK}.}
    \label{fig:c2c-combined}
\end{figure}

\noindent
This second example explicitly tests the regulated pipeline on an ill-posed Cauchy-to-Cauchy transformation where
the target smearing width $\varepsilon_2$ is decreased relative to the input Cauchy width $\varepsilon_1$.
Due to the divergence of \cref{eq:c2c}, the target kernel is replaced by a Tikhonov-regulated approximant, following Sec.~\ref{subsec:RK_bounds_regulated}.
For each target energy $\omega'$, the coefficient vector $g_{\regulator}(\omega')$ introduced in \cref{eq:g_reg1,eq:g_reg2} defines a regulated Cauchy target kernel, the propagated half-width $\Delta_{\rm prop}(\regulator)$, the kernel-mismatch bias $\Delta_{\rm sys}(\regulator)$, and the total certified half-width $\Delta_{\rm tot}(\regulator)$ of \cref{eq:total_systematic_error}.
The regulator is chosen as the minimizer in \cref{eq:optimized_regulator_T}, i.e.
\begin{equation}
    \regulator^\star \in \argmin_{\regulator} \Delta_{\rm tot}(\regulator) \,.
\end{equation}
Both RK and positivity-optimized bounds can then be applied to the regulated target observable, with the final interval enlarged to include the certified kernel-mismatch bias.
\Cref{fig:c2c-combined} shows the regulator scan, the certified kernel approximation at the chosen regulator, and the resulting reconstructions across target energies.

As illustrated in Fig.~\ref{fig:c2c-optimization}, the two error components exhibit competing
behaviors. As $\regulator \to 0$, the regulator is removed and the systematic approximation bias vanishes,
but the propagated half-width grows due to the highly oscillatory transition weights, $g_{\regulator}$. Conversely,
a large $\regulator$ heavily damps their fluctuations, reducing error propagation but failing to
accurately reconstruct the sharper target Cauchy peak.
As anticipated above, for a given target energy $\omega'$, the regulator
$\regulator=\regulator^\star$ is chosen by minimizing the total certified half-width, cf. \cref{eq:total_systematic_error}, as shown in \cref{fig:c2c-optimization}.
Instead, \cref{fig:c2c-comparison}
shows the computation of the approximate kernel at the optimal $\regulator^\star$, alongside the
corresponding upper and lower bounds from the SIP method.
The bottom panel confirms the
validity of the certificate on the kernel-mismatch contribution,
\cref{eq:bound_rho_kappa_from_data}.
As above, the choice of input smearing width has been optimized to minimize the overall error.
The certified bias remains rigorous, with the largest error occurring around the peaks of the spectral functions, which unsurprisingly are the hardest to reconstruct.
As expected, the total error grows dramatically if the target smearing is reduced for fixed input smearing.
This growth depends not only on the ratio $\varepsilon_2/\varepsilon_1$, but also on the absolute size of the widths.

It is also useful to consider limiting cases for the input smearing width $\varepsilon_1$ at fixed target width $\varepsilon_2$, in analogy with \cref{fig:stability_analysis}.
As $\varepsilon_1$ approaches the target resolution $\varepsilon_2$, the kernel transformation becomes
progressively closer to an identity map; the transition coefficients remain highly localized and stable,
preventing the inflation of the total error.
However, obtaining such input data is difficult, since fine resolutions typically involve large uncertainties (scaling as $\Delta\rho_{\varepsilon_1} \propto 1/\varepsilon_1^2$ in the present example).
This behavior, of course, is a manifestation of the original ill-posed inverse Laplace transform.
Conversely, for very wide input data (large $\varepsilon_1$), the data become extremely
precise, but the numerical deconvolution requires violently oscillating transition coefficients that exponentially amplify the propagated half-width.
The same stability/resolution trade-off therefore reappears in the regulated setting, with a slower growth of RK bounds as $\varepsilon_1$ increases.

For the purposes of this example, focused on the rigorous Cauchy-to-Cauchy sharpening,
the above analysis is instead repeated across different target energies\footnote{
    Although similar results are obtained by minimizing $\Delta_\text{tot}$ at each value of $\omega'$, we numerically find that separately optimizing $\xi$ to identify the optimal upper and lower RK bounds, as in \cref{eq:rho_kappa_pm_rig_RK} and as done for the SIP case, stabilizes the search for the optimal regulator. 
    Hence, while the symmetric optimization is shown for \cref{fig:c2c-optimization}, the asymmetric strategy is employed for \cref{fig:c2c-reconstructions}.
}, at fixed $\varepsilon_1$
and $\varepsilon_2=\varepsilon_1/2$.
The results are shown in \cref{fig:c2c-reconstructions}.
In this controlled example, the positivity-optimized formulation gives tighter bounds, albeit at the price of a more involved numerical procedure.
The input errors, which are of order $10\%$, are unavoidably
inflated and often more than doubled.
This example illustrates both the difficulties of the
deconvolution, for example extracting the spectral function smeared with a more localized Cauchy kernel than
that of the input data, and the success of the numerical procedures outlined in this paper in estimating
the resulting certified bounds.

\subsection{L\'evy-mixture kernel transformation (Gaussian to Cauchy)\label{sec:num-levy}}
\noindent
As an additional numerical example, this Section presents a numerical test of the Gaussian-to-Cauchy transformation using the L\'evy mixture in \cref{eq:Levy}.
The purpose of the example is to illustrate how a change of perspective---moving from input data with a fixed Gaussian width to a continuum of Gaussian widths---removes the obstruction in the fixed-width case described in Sec.~\ref{subsec:analytic_kernel_transformations}.
The example adopts the same model spectral function as in \cref{eq:toy_spectral_function}, but modifies the model bounds relative to \cref{eq:toy_bounds} in order to account for their quadratic divergence as $\sigma\to0$.
The following modification is applied:
\begin{gather}
    \rho_{\sigma+}^{\mathtt g}(\omega) = \rho_\sigma^{\mathtt g}(\omega) + \left(\frac{\sigma_+}{\sigma}\right)^2 \,, \\
    \rho_{\sigma-}^{\mathtt g}(\omega) = \rho_\sigma^{\mathtt g}(\omega)
    \left[ 1+\left(\frac{\sigma_-}{\sigma}\right)^2 \right]^{-1} \,,
\end{gather}
with the values $\sigma_+=0.55$ and $\sigma_-=0.60$ taken for concreteness.
The L{\'e}vy transition kernel is plotted in \cref{fig:gaussian_to_cauchy_kernel}.
\begin{figure}[t!]
    \centering
\subfloat[Transition kernel and input smeared spectral function for Gaussian-to-Cauchy kernel transformations.
\label{fig:RK_bounds_Levy_a}]{%
        \includegraphics[width=0.48\textwidth]{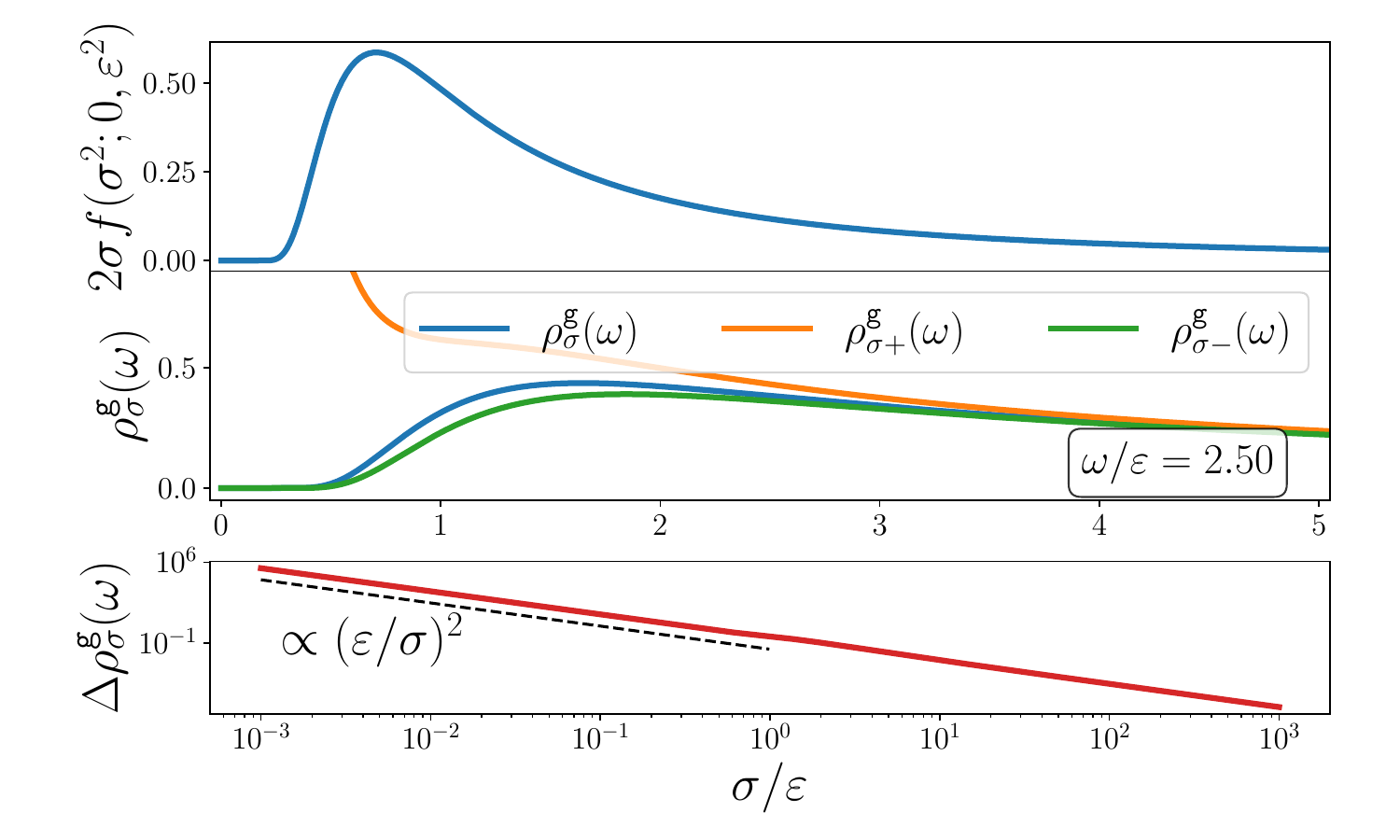}%
}
    \hfill
\subfloat[Integrand and derivation of RK bounds for Gaussian-to-Cauchy kernel transformations.
\label{fig:RK_bounds_Levy_b}]{%
        \includegraphics[width=0.48\textwidth]{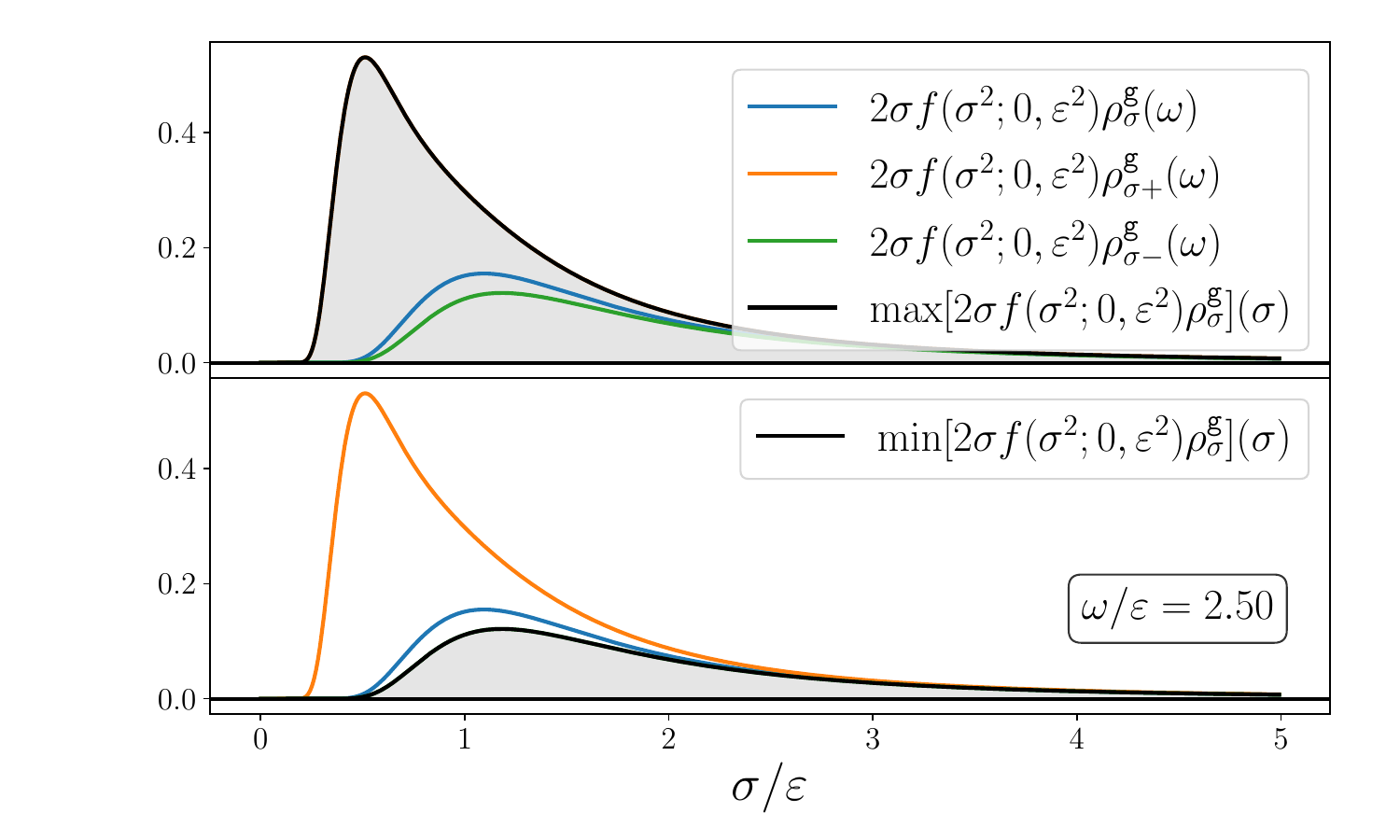}%
} \caption{RK bounds with the positive L\'evy kernel, see \cref{eq:Levy}.
\Cref{fig:RK_bounds_Levy_a} shows the individual contributions of the transition kernel, the smeared spectral function and its bounds, and the divergence of the latter as a function of $\sigma/\varepsilon$, where $\sigma$ labels the smearing width of the input Gaussian and $\varepsilon$ that of the output Cauchy smearing functions.
\Cref{fig:RK_bounds_Levy_b} instead shows how RK bounds are propagated.
Note that this is straightforward given the positivity of the L\'evy kernel.}
    \label{fig:RK_bounds_Levy}
\end{figure}
\begin{figure}[t!]
    \centering
    \includegraphics[width=0.99\linewidth]{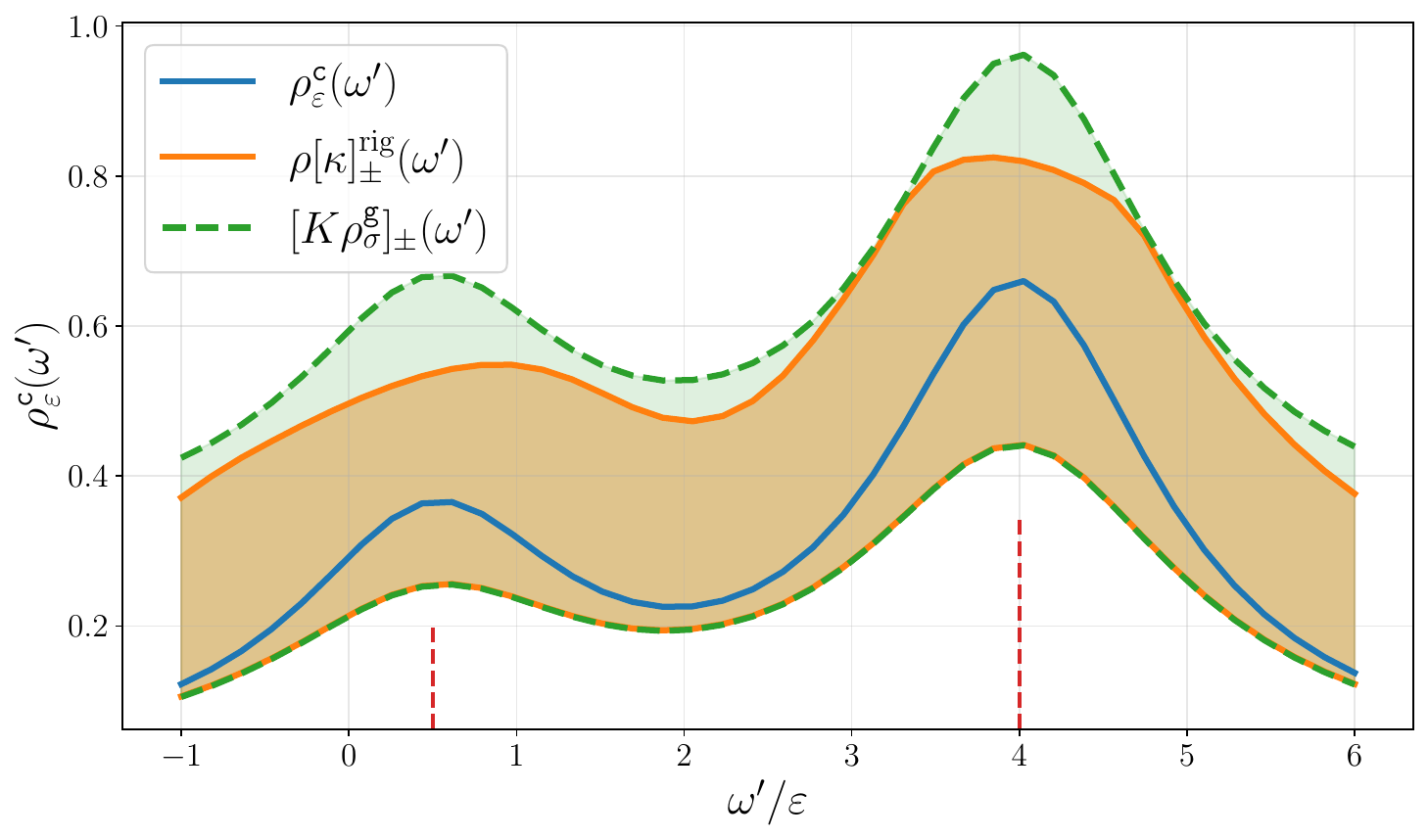}
\caption{Gaussian-to-Cauchy reconstructions at different values of the center $\omega'$ of the target Cauchy kernel with a fixed smearing width $\varepsilon$.
The figure shows results for RK and SIP bounds in green and orange, respectively, both computed from the Gaussian-smeared input data.}
    \label{fig:reconstructions_Levy}
\end{figure}

\Cref{fig:RK_bounds_Levy} shows the construction of RK bounds from propagating the known bounds on the input Gaussian-smeared spectral function to the Cauchy-smeared spectral function.
In particular, the upper and lower bounds correspond to the two areas highlighted in \cref{fig:RK_bounds_Levy_b}.
The positivity of the L{\'e}vy kernel and that of its associated integral operator allow for a more intuitive interpretation of the corresponding RK bounds.
Note also that the quadratic divergence of the bounds is suppressed by the exponential decay of the L\'evy kernel.

An important distinction compared to the Cauchy-to-Gaussian transformation is the lack of a tunable external parameter analogous to $\varepsilon$ in \cref{fig:stability_analysis}.
In this case, the mapping involves a linear combination of all Gaussian widths as in \cref{fig:RK_bounds_Levy}, with the target center $\omega'$ held fixed and equal to the input center.
\Cref{fig:reconstructions_Levy} shows the behavior of the corresponding reconstructions as a function of $\omega'$.
Both the RK and SIP bounds are included, as discussed in Sec.~\ref{subsec:positivity_bounds}.
While the positivity of the L{\'e}vy kernel guarantees that RK bounds are straightforward to compute without negative-weight cancellations, they are not necessarily optimal, as the RK construction allows the spectral density to saturate the input bounds independently across all energies.
The SIP approach naturally accounts for the global constraints of the spectral density and yields tighter upper bounds than the RK integration in this example, as seen in \cref{fig:RK_bounds_Levy}.
To compute the upper SIP bound numerically, the coefficients $\lambda_i$ in the optimization procedure in \cref{eq:dual-upper} must be held positive, mimicking the same property of the exact transition kernels.
While this additional constraint is neither required nor appropriate for the other transitions examined in this work, it is needed here for the computation of the upper bounds and the SIP certificates in \cref{eq:certification_SIP}.
It is also useful to compare the RK and SIP bounds.
In this example, the SIP upper bound improves on the RK upper bound, while the lower SIP and RK bounds are consistent.

\subsection{Spectral reconstructions as kernel transformations}

\begin{figure}[t!]
    \centering
\subfloat[Scan over $\omega'$ at fixed $\sigma$ and $N_t=24$.
\label{fig:ilt}]{%
        \includegraphics[width=0.48\textwidth]{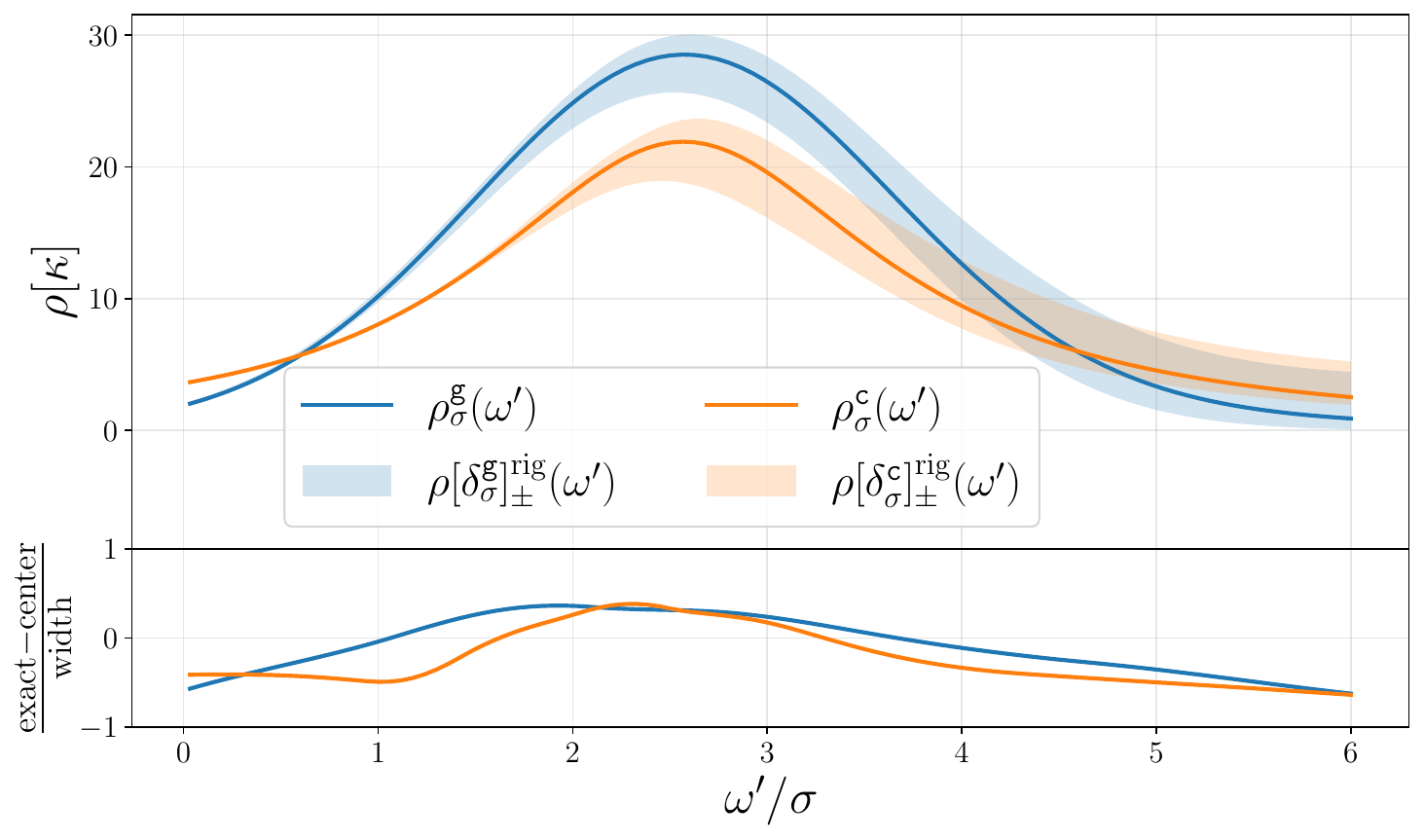}%
}
    \hfill
\subfloat[Scan over $N_t$ at fixed $\omega'$ and $\sigma$.
\label{fig:ilt_Nt}]{%
        \includegraphics[width=0.48\textwidth]{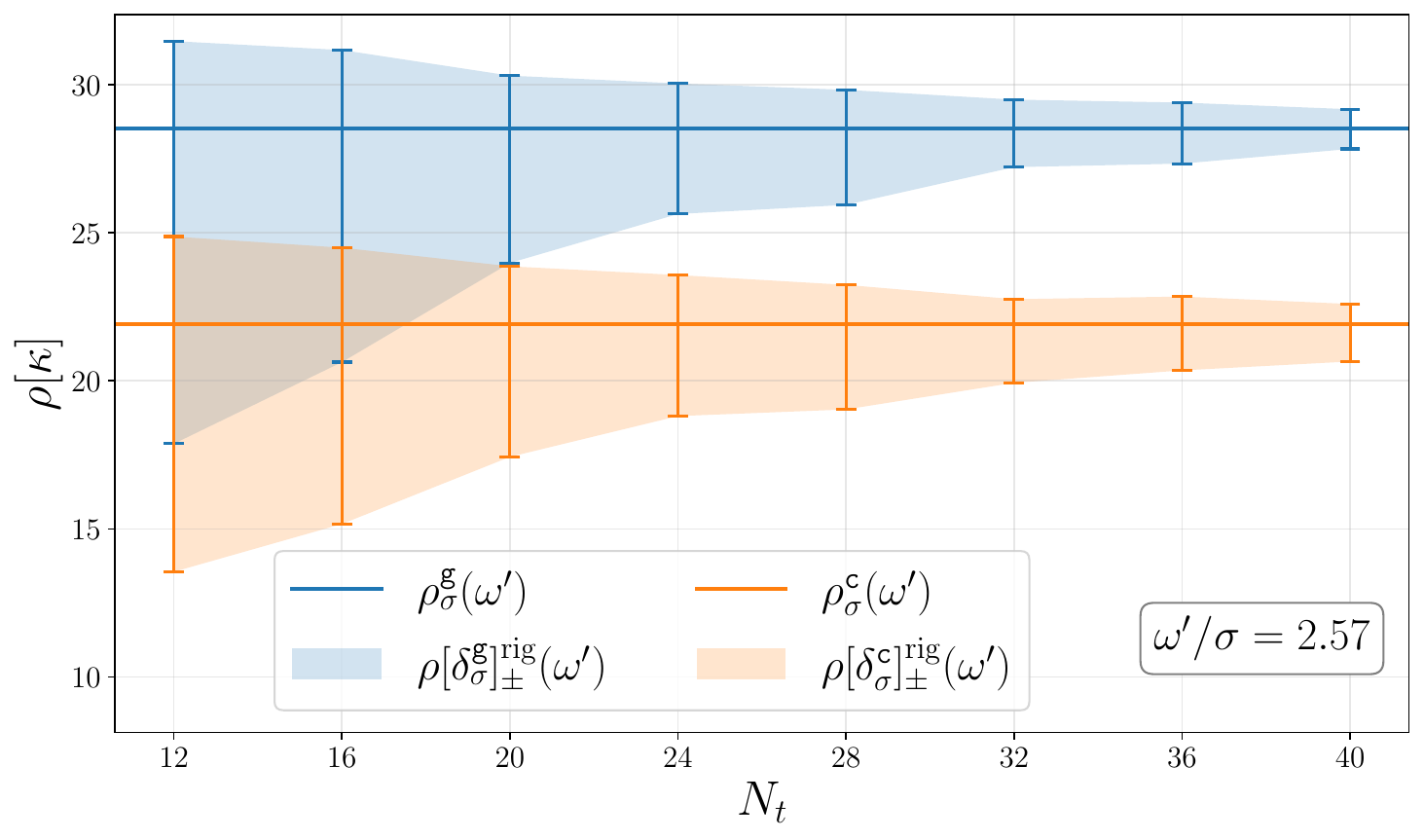}%
} \caption{Exponential-to-Gaussian and exponential-to-Cauchy reconstructions as a function of the target energy $\omega'$ with a common smearing width $\sigma=0.300$.
The figure shows results for SIP bounds, both derived from the Euclidean correlation function as input data. 
The number $N_t$ of available time-slices is fixed by requiring a constant $30\%$ noise-to-signal at the last evaluated time slice. \Cref{fig:ilt} shows results at fixed $N_t=24$ and $\sigma$, across different values of $\omega'$. \Cref{fig:ilt_Nt} instead displays results for different values of $N_t$, at fixed $\omega'$ and $\sigma$.}
    \label{fig:ilt_all}
\end{figure}

\noindent
This final numerical example illustrates how kernel transformations might be adopted for practical reconstructions to establish bounds on smeared spectral densities from Euclidean-time correlation functions.

For this purpose, a different spectral function model is adopted in order to mimic the finite-volume spectral profile of a resonance.
The Euclidean-time correlator input data is generated from a multivariate Gaussian distribution with central values\footnote{All quantities are given in units of the lattice spacing $a=1$.}
\begin{equation}
    C_t = \sum_{k=0}^{N_{\rm states}-1} |Z_k|^2 e^{-E_k t} \,, \quad
    t = 1, \dots, N_t
\end{equation}
with amplitudes and energies
\begin{equation}
    |Z_k|^2 = z_k^2 \delta_\Gamma^\mathtt{c}(M,E_k) \,, \quad
    z_k = \begin{cases}
        1/5 \,, & k \text{ even} \,, \\
        1   \,, & k \text{ odd}  \,,
    \end{cases}
\end{equation}
\begin{equation}
    E_k = m + (k+1) \, \Delta E \,.
\end{equation}
The covariance matrix is taken to have the form
\begin{equation}
    \Sigma_{t,t'} = \sigma_0^2 \, [s+(1-s)\delta_{t,t'}] e^{-\gamma|t-t'|} \, e^{\Delta m (t+t')} C_t C_{t'} \,.
\end{equation}
In this model, $\Delta m=E_0/2-m$ controls the signal-to-noise degradation, while $m=0.135$ represents a nonzero mass gap; $\gamma^{-1}=1.3$ measures correlation, $s=0.5$ is a shrinkage parameter. The overall scale $\sigma_0$ is set to match a $30\%$ noise-to-signal ratio at the last evaluated time slice, $N_t$. The energy spacing is fixed to $\Delta E = 0.020$, while the number of states is $N_{\rm states} = 100$. The resonance is localized at $M=0.770$ with a width of $\Gamma=0.075$.

\Cref{fig:ilt_all} illustrates the reconstruction of Gaussian- and Cauchy-smeared spectral functions from these noisy input data.
The solutions are obtained by solving and certifying the semi-infinite program defined in \cref{eq:primal-upper,eq:dual-upper}, suitably adjusted to incorporate the compatibility with noisy input data in place of pointwise bounds.\footnote{ \label{foot:exchange}
    We employ an exchange algorithm that iteratively solves the SIP on an increasingly dense energy grid via primal-dual interior-point methods (like \textsc{clarabel}~\cite{clarabel}) up to a predefined numerical tolerance, and subsequently use \cref{eq:rho_kappa_pm_rig} to certify the final result against the small but inevitable residual violations, as commented in Sec.~\ref{subsec:positivity_certification}. 
} 
In particular, the primal problem becomes~\cite{Lawrence:2024hjm,Abbott:2025snz,Mutzel:2026vyw}
\begin{equation}\label{eq:primal-upper-noisy}
\begin{aligned}
    &\rho[\kappa]_+^\star
    = \max_{\rho} \quad \rho[\kappa]
    \\
    & \qquad \text{such that} \,\, \Vert C_t-\rho[\phi_t] \Vert_\Sigma \leq \chi_0^2 \,, \\
    & \qquad \text{and} \,\, \rho(\omega) \geq 0 \quad \forall \omega \in [0, \infty) \,,
\end{aligned}
\end{equation}
where $\Vert x \Vert_\Sigma \equiv x^T \Sigma^{-1} x$ and $\rho[\phi_t]$ is defined in \cref{eq:inverse_problem}.
The associated dual problem can be derived as in Ref.~\cite{Abbott:2026wdw} and reads
\begin{equation}\label{eq:dual-upper-noisy}
\begin{aligned}
        &\rho[\kappa]_+^\star = \max_{g \in \mathtt{R}^{N_t}} g^T C - \chi_0 \sqrt{g^T \Sigma g} \\
        & \qquad \text{such that} \,\, \sum_{t=1}^{N_t} g_t \phi_t(\omega) \geq \kappa(\omega) \quad \forall \omega \in [0,\infty) \,.
\end{aligned}
\end{equation}
The lower panel of \cref{fig:ilt} confirms visually the validity of the rigorous bounds for both smearing kernels across different energies, at fixed $N_t=24$.
Note that \cref{eq:primal-upper-noisy} is equivalent to the primal problem considered in Refs.~\cite{Lawrence:2024hjm,Abbott:2026wdw,Mutzel:2026vyw}.
The difference in the present work is that the problem is first solved approximately, and then corrected with an explicit certification step.$^{\ref{foot:exchange}}$

\Cref{fig:ilt_Nt} shows the improvement of bounds as the number of Euclidean times is increased from 12 to 40 for fixed values of $\omega'$, located at the peak, and $\sigma$.
To mirror a typical analysis strategy, the number of inputs is increased while holding the noise-to-signal ratio fixed at $30\%$ on the last time slice. 
As expected, the bounds improve with the inclusion of additional statistically significant data.

Solving \cref{eq:primal-upper-noisy} requires a choice of the statistical cutoff parameter $\chi_0^2$.
For simplicity, and since the exact values are known in this example, an \emph{ad hoc} value of $\chi_0^2/N_t\approx 0.8$ was chosen to match the $\chi^2$ between the noisy and exact correlator values.
As emphasized in Refs.~\cite{Lawrence:2024hjm,Abbott:2026wdw,Mutzel:2026vyw}, this choice has a statistical interpretation.
A rigorous statistical characterization of tight confidence intervals associated with the choice of the parameter $\chi_0^2$ remains an open research problem.
In phenomenological applications, care must be taken in choosing $\chi_0^2$ appropriately.

In practical applications, the dual condition can only be enforced on a suitably large, but finite, energy domain. 
When this cutoff is not fixed by underlying kinematical constraints, the rapid suppression of high-energy contributions by the input smearing functions motivate this truncation.
The effects of this truncation were verified to be numerically negligible for the energy scales considered in \cref{fig:ilt}.

It is important to note that this is not the only viable approach to spectral reconstruction. 
In fact, combining the technology developed in this work with that of the causal bootstrap developed in Ref.~\cite{Abbott:2026wdw} suggests at least two avenues for future work.
To motivate these alternatives, it is useful to highlight that a key requirement for the causal bootstrap reduction to a finite-dimensional semidefinite program is that the target kernel must be a (piecewise) polynomial or rational function in the relevant variable.
For example, since moment problems are naturally given in terms of the relevant variable $\lambda=e^{-\omega}$, it is clear that this restriction formally precludes the exact computation of two-sided bounds for Gaussian or Cauchy smearing kernels in the energy domain starting from Euclidean-time data.
This can be seen from the SIP perspective in terms of substantially different UV behavior of input and target smearing kernels. More explicitly, the linear combinations of decaying exponentials forming the candidate bounding smearing kernel in \cref{eq:kappa_lambda} can never indefinitely bound a Cauchy kernel from above or a Gaussian kernel from below.
On the other hand, as shown in Ref.~\cite{Abbott:2026wdw}, the $\lambda$-domain Cauchy kernel $\delta_\varepsilon^\mathtt{c}(\lambda)$ \emph{is} accessible.
Note that, when explicitly rewritten in energy space and by properly including the suitable change of measure, the resulting smearing kernel takes the form of \cref{eq:Cauchy-like} with $t_0=1$. Therefore, bounds on this quantity are also accessible through \cref{eq:bound_rho_kappa_from_data}.

The first hybrid strategy is to use a polynomial approximation of the desired smearing kernel on a given, finite energy domain, for which the causal bootstrap gives exact, tight bounds.
A similar scheme has been advocated in Ref.~\cite{Mutzel:2026vyw}, which computes bounds through a sequence of ``relaxations" constructed to lie above or below the target kernel.
Regardless of the precision of the approximation, the inevitable residual systematic error can be bounded via \cref{eq:bound_rho_kappa_from_data}, which is novel to the present work.

A second possibility is to begin by extracting rigorous bounds for some convenient intermediate smeared spectral function which is directly accessible using the causal bootstrap, for example the Cauchy-like smearing kernel in \cref{eq:Cauchy-like}.
Then, in a subsequent step, one could transform to smearing kernels of direct phenomenological relevance using the techniques developed in the manuscript.

%% file: sec_conclusions.tex
\noindent
This work has developed a framework for constructing maps between smeared spectral functions defined with different smearing kernels.
Although the methods can be applied directly to Euclidean-time inputs, the main focus has been on the kernel-transformation problem: given continuous energy-smeared input observables $\rho[\phi_\alpha]$, and possibly rigorous pointwise bounds on them, determine a target smeared observable $\rho[\kappa]$ without returning to the original inverse problem.
Exact transformations are controlled by the analytic and Fourier structure of the input and target kernels.
When an exact transformation is not available, the target kernel can be replaced by a regulated approximant, and the resulting bias from kernel mismatch can be sharply quantified from the available input data.
Pointwise rigorous bounds then propagate through the transformation either via a general sign-splitting argument or, where applicable, through optimized bounds that leverage positivity of the underlying spectral function.
Extensions to statistical error propagation can be naturally incorporated as well.

This framework supports the analysis strategy emphasized in the introduction.
Rather than laboriously returning to the original inverse problem in terms of a finite Euclidean-time basis for every desired inclusive observables, one instead has the freedom to target the most convenient smeared spectral information:~\mbox{$\rho[\phi_t=e^{-\omega t}] \to \rho[\phi_\alpha]$}.
Exact or certified regulated kernel transformations can then transport that information to other smeared or weighted observables.
The resulting transformation path can be chosen to minimize the total certified uncertainty, including propagated input uncertainty, loss of correlations arising from intermediate transformations, and (where relevant) any systematic contribution from kernel mismatch.

\begin{table}[t!]
    \centering
    \caption{
    Summary of kernel transformations.
    The first block describes transformations from the discrete Euclidean time basis, which provide important motivation but are not directly considered in the present work.
    The second and third blocks describe the transformations between energy-smeared observables that are the focus of the present work.
    The final column summarizes the conditions under which an unregulated bijection $K^{\kappa\leftarrow\phi}$ exists.
    \label{tab:kernels}
    }
    \begin{tabular}{l l | l l }
        \toprule
        \toprule
        \textbf{$\phi_\alpha(\omega)$} & $\alpha$ domain & \textbf{$\kappa(\omega)$} & $K^{\kappa \leftarrow \phi}$ exists? \\
        \midrule
        $e^{-\omega \alpha}$ & $\alpha \in \N$
          & $a_\mu^{\mathrm{HVP}}$                              & Yes, Ref.~\cite{Bernecker:2011gh} \\
        & & $\Delta\alpha_{\mathrm{EM}}$                        & Yes, Ref.~\cite{Bernecker:2011gh} \\
        & & $\delta_\varepsilon^c(\omega,\omega')$              & No \\
        & & $\delta_\varepsilon^g(\omega,\omega')$              & No \\
        \midrule
        $\delta_\varepsilon^{\mathtt c}(\alpha,\omega)$ & $\alpha \in \R$
          & $\delta_{\varepsilon'}^{\mathtt c}(\omega,\omega')$ & Yes, if $\varepsilon' \geq \varepsilon$ \\
        & & $\delta_{\varepsilon'}^{\mathtt g}(\omega,\omega')$ & Yes, for all $\varepsilon'$ \\
        & & $\Pi_{\varepsilon'}$                      & Yes, if $\varepsilon' \geq \varepsilon$ \\
        & & $\Gamma_\sigma^{\mathtt g}$                              & Yes\\
        & & $\Gamma_{\varepsilon'}^{\mathtt c}$                      & Yes, if $\varepsilon' \geq \varepsilon$ \\
        [0.5em]

        $\delta_\sigma^{\mathtt g}(\alpha,\omega)$ & $\alpha \in \R$
          & $\delta_{\sigma'}^{\mathtt g}(\omega,\omega')$                  & Yes, if $\sigma' \geq \sigma$ \\
        & & $\delta_\varepsilon^{\mathtt c}(\omega,\omega')$    & No \\
        & & $\Gamma_{\sigma'}^{\mathtt g}$                           & Yes, if $\sigma' \geq \sigma$\\
        & & $\Gamma_{\sigma'}^{\mathtt c}$                           & No\\
        \midrule

        $\delta_\alpha^{\mathtt g}(\omega,\omega')$ & $\alpha \in \R^+$
          & $\delta_\varepsilon^{\mathtt c}(\omega,\omega')$ & Yes, for all $\varepsilon > 0$ \\
        $\delta_\alpha^{\mathtt c}(\omega,\omega')$ & $\alpha \in \R^+$
          & $\delta_{\sigma}^{\mathtt g}(\omega,\omega')$       & No \\
        \bottomrule
        \bottomrule
    \end{tabular}
\end{table}

\Cref{tab:kernels} summarizes the main cases considered in the previous Sections.
The first lesson is that exact transformations are available when the input family and target kernel have compatible analytic structure.
In particular, the table is organized into three blocks, corresponding to different classes of kernel transformations. 
The first block describes known examples of inverse problems from Euclidean-time correlators. 
The second block includes the widening semigroups for Gaussian and Cauchy kernels and, more importantly for the present work, the Cauchy-to-Gaussian map used as the main exact example for rigorous bound propagation.
The third block contains the case of the L\'evy mixture of different smearings at fixed energy, which allows a Gaussian-to-Cauchy transformation.

The bound-propagation methods are complementary.
RK bounds are direct, inexpensive, and apply to any signed transition kernel once pointwise input bounds have been specified.
Positivity-optimized bounds enforce the physical requirement of the positivity of the underlying spectral density,
and can therefore yield tighter intervals at the price of a numerical optimization and a certification step.
The numerical examples illustrate both methods in controlled settings with a toy model mimicking realistic lattice-data analysis.

Regulated transformations extend the toolkit to cases where exact maps fail or where a finite-dimensional approximation is used in practice.
A key novel observation in the present work is that bias can be bounded from the knowledge of input data only, and the regulator can be tuned to produce optimized, certified bounds.
The Cauchy-to-Cauchy sharpening example illustrates this procedure and shows explicitly how the cost of sharpening enters through enlarged, rigorously certified intervals.

Looking forward, the propagation of absolute pointwise bounds introduced in this work is particularly well suited for interfacing with recent developments that yield strict analytic domains (\emph{Wertevorr\"ate}) for Cauchy-smeared spectral functions~\cite{Bergamaschi:2023xzx,Abbott:2025snz,Fields:2025glg,Abbott:2026wdw}.
Kernel transformations provide the mathematical bridge from those Cauchy-smeared constraints to physical, kinematically weighted observables.
Moreover, the certification of semi-infinite solutions offers a complementary strategy to those presented in Refs.~\cite{Abbott:2026wdw,Lawrence:2024hjm,Mutzel:2026vyw}, which compute tight bounds for other kernels.
Another interesting application may lie in propagating the finite-volume systematic uncertainties, recently computed for Cauchy-smeared spectral functions in Ref.~\cite{Bresciani:2026kjv}, to other smearing kernels.
Also related to systematic uncertainties, bounds arising from kernel transformations offer a broad range of self-consistency conditions for different smeared spectral quantities, which can be used to verify the internal consistency of a calculation.
To retain the benefit of kernel transformations without unnecessarily inflating the associated errors, correlation on bounds on intermediate smeared spectral observables must be properly propagated.
The natural next step is to apply these methods in practical LQCD calculations.

%% file: app_normalization.tex
\section{\MakeUppercase{Normalization of transition kernels}}\label{app:normalization}
\noindent
This short Appendix records the normalization convention used in Sec.~\ref{subsec:general-realization}.
Given a pair of input and target smearing kernels, $\kappa^{\mathtt x}_\varepsilon$ and $\kappa^{\mathtt x'}_\sigma$ respectively, the corresponding transition kernel $K$ satisfies
\begin{equation}
    \kappa^{\mathtt x'}_\sigma(\omega,\omega') = \int_{\R} d\omega'' \,
    K_{\sigma\leftarrow\varepsilon}^{\mathtt x'\leftarrow\mathtt x}(\omega,\omega'')
    \kappa^{\mathtt x}_\varepsilon(\omega'',\omega') \,.
\end{equation}
In this Appendix, conditions that relate the normalization of the input, target, and transition kernels are reviewed.

When both input and target kernels are normalized with respect to integration of the second argument, i.e.
\begin{equation}
    \int_{\R} d\omega' \, \kappa^{\mathtt x}_\varepsilon(\omega,\omega') =
    \int_{\R} d\omega' \, \kappa^{\mathtt x'}_\sigma(\omega,\omega') = 1 \,,
\end{equation}
then
\begin{align}
    1 &= \int_{\R} d\omega' \, \kappa^{\mathtt x'}_\sigma(\omega,\omega') \\
    &= \int_{\R} d\omega' \, \int_{\R} d\omega'' \,
    K_{\sigma\leftarrow\varepsilon}^{\mathtt x'\leftarrow\mathtt x}(\omega,\omega'')
    \kappa^{\mathtt x}_\varepsilon(\omega'',\omega') \\
    &= \int_{\R} d\omega'' \,
    K_{\sigma\leftarrow\varepsilon}^{\mathtt x'\leftarrow\mathtt x}(\omega,\omega'')
\end{align}
that is, the transition kernel is also normalized with respect to integration of the second argument.

If both the input and target kernels are normalized with respect to integration of the first argument, i.e., if
\begin{equation}
    \int_{\R} d\omega \, \kappa^{\mathtt x}_\varepsilon(\omega,\omega') =
    \int_{\R} d\omega \, \kappa^{\mathtt x'}_\sigma(\omega,\omega') = 1
\end{equation}
then
\begin{align}
    1 &= \int_{\R} d\omega'' \, \kappa_\varepsilon^{\mathtt x}(\omega'',\omega') \\
    &= \int_{\R} d\omega \, \kappa^{\mathtt x'}_\sigma(\omega,\omega') \\
    &= \int_{\R} d\omega'' \, \kappa^{\mathtt x}_\varepsilon(\omega'',\omega')
    \cdot \left[ \int_{\R} d\omega \,
    K_{\sigma\leftarrow\varepsilon}^{\mathtt x'\leftarrow\mathtt x}(\omega,\omega'') \right] \,.
    \label{eq:norm_second_arg}
\end{align}
By equating the first and last equations and supposing that
\begin{equation}
    \int_{\R} d\omega'' \, \kappa_\varepsilon^{\mathtt x}(\omega'',\omega') f(\omega'')
    = 0 \quad \text{iff} \quad f(\omega'') = 0 \,,
\end{equation}
the transition kernel is also normalized with respect to integration of the first argument.
This condition is automatically satisfied if the kernels are symmetric.

The above equations simplify substantially for translationally invariant kernels, where the transition kernel is also translationally invariant. The integral in \cref{eq:norm_second_arg} factorizes, guaranteeing the normalization of the transition kernel provided that of the input and target smearing kernels.

%% file: app_rk.tex
\section{\MakeUppercase{Riesz--Kantorovich bounds from abstract operator theory}}\label{app:RK_bounds}
\noindent
This Appendix gives the operator-theory support for the RK sign-splitting formulas used in Sec.~\ref{subsec:RK_bounds}.
It considers the propagation of pointwise uncertainties under convolution.
In particular, an operator-theoretic argument is presented to translate bounds on smeared spectral quantities between different kernels.

Let $f$ and $K$ be well-behaved functions on the real line.
Consider the linear functional
\begin{align}\label{eq:general_linear_integral_operator}
    Tf= \int_{\R} dx \, K(x) f(x) \equiv F
\end{align}
wherever the integral is well defined.
Suppose that pointwise non-negative bounds for $f$ are given
\begin{align}
    0 \leq f_-(x) \leq f(x) \leq f_+(x) \,.
\end{align}
The theory of order-bounded linear operators provides an abstract way to translate bounds on $f$ into bounds on $F$.
A self-contained presentation of the relevant operator theory can be found in Chapter 1 of the textbook Ref.~\cite{AliprantisBurkinshaw2006}, together with references to the original mathematical literature.

The discussion begins with the notion of a positive operator.
An operator $T$ between ordered vector spaces is said to be positive if $Tx \geq 0$ for all $x \geq 0$, and monotone whenever $x\geq y$ implies $Tx \geq Ty$.
Suppose that $T$ is a positive, monotone operator.\footnote{A familiar example of a nonmonotone operator is that induced by the (unregulated) principal-value distribution,
    \begin{equation}
        Tf = {\rm P.V.}\int dx\, \frac{f(x)}{x} \,.
    \end{equation}
}
Acting with the operator $T$ on the bounds for $f$ gives the obvious bounds for $F=Tf$,
\begin{align}
    F_- \leq F \leq F_+
\end{align}
where $F_-$ is the greatest lower bound ($\inf$) and $F_+$ is the least upper bound ($\sup$)
\begin{align}
    F_-
    &= \inf_{f_- \leq f \leq f_+} Tf = Tf_- \\
    F_+
    &= \sup_{f_- \leq f \leq f_+} Tf = Tf_+ \,.
\end{align}

These bounds can be extended to the case when $T$ is a difference of positive monotone operators, i.e., the case where the kernel $K(x)$ in \cref{eq:general_linear_integral_operator} takes on both positive and negative values.
To see this, decompose the kernel $K: \R\to \R$ as the difference
\begin{align}
    K  = K^+ - K^-
\end{align}
where
\begin{align}
    K^\pm &: \R \to \R\\
    K^\pm(x) &= \max\{ \pm K(x), 0\} \,.
\end{align}
Pointwise bounds on the integrand $K(x) f(x)$ are given by
\begin{align}
    \nonumber
    \inf_{f_- \leq f \leq f_+}
    K(x) f(x)
    &=\begin{cases}
        K(x) f_-(x), & K(x) \geq 0 \\
        K(x) f_+(x), & K(x) < 0
    \end{cases}
    \\
    &=
    (K^+ f_- - K^- f_+)(x)\\
    &\equiv \inf (K f)(x) \,,
    \\
    \nonumber
    \sup_{f_- \leq f \leq f_+}
    K(x) f(x)
    &=\begin{cases}
        K(x) f_+(x), & K(x) \geq 0 \\
        K(x) f_-(x), & K(x) < 0 \,.
    \end{cases}
    \\
    \label{eq:upper_bound_breakdown}
    &= (K^+ f_+ - K^- f_-)(x)
    \\
    &\equiv \sup (K f)(x) \,.
\end{align}
In each case, the first equalities handle the bounds according to whether the kernel is positive or negative.
The second equalities express the bounds in terms of $K=K^+ - K^-$.
Finally, the third lines establish compact notation for the integrands.
With the help of this notation, bounds on the integrals take the following simple form:
\begin{align}
    \label{eq:lower_bound_Krho_appendix}
    F_- &= \int_{\mathbb R} dx \, \inf(K f)(x) \,,
    \\
    \label{eq:upper_bound_Krho_appendix}
    F_+ &= \int_{\mathbb R} dx \, \sup(K f)(x) \,.
\end{align}
Such bounds are referred to as RK bounds in this work.
These are the abstract generalizations of \cref{eq:upper_bound_Krho,eq:lower_bound_Krho} in the main text, the only difference being more cautious treatment of $\min/\inf$ and $\max/\sup$.
This distinction plays no role in the present applications.

%% file: app_sip_dual.tex
\section{\MakeUppercase{Derivation of the Semi-Infinite Programming Dual}}\label{app:sip_dual}
\noindent
This Appendix gives the primal-dual derivation supporting the positivity-optimized bounds in Sec.~\ref{subsec:positivity_bounds}.
The unconstrained dual problem, \cref{eq:dual-upper}, is derived from the semi-infinite primal problem, \cref{eq:primal-upper}.
The focus will be on the upper bound; the lower bound follows from an identical procedure by negating the objective function.

The primal problem seeks to maximize the functional $\rho[\kappa] = \int_0^\infty d\omega \, \kappa(\omega)\rho(\omega)$ subject to the spectral positivity constraint $\rho(\omega) \ge 0$ and a discrete set of $L_\infty$ box constraints on the input data
\begin{equation}
    \overline\rho_i - \Delta\rho_i \le \int_0^\infty d\omega \, \phi_i(\omega)\rho(\omega)
    \le \overline\rho_i + \Delta\rho_i \,,
\end{equation}
where $i$ indexes the input kernels and we adopt the shorthand notation $\overline\rho_i \equiv \overline\rho[\phi_{\alpha_i}]$ and $\Delta\rho_i \equiv \Delta\rho[\phi_{\alpha_i}]$.
The solution of the constrained problem is obtained by the method of Lagrange multipliers.

To construct the Lagrangian, two sets of non-negative dual variables are introduced: $\nu(\omega) \geq 0$ for all $\omega$, associated with the spectral positivity constraint, and $\lambda_i^+ \ge 0$ and $\lambda_i^- \ge 0$, associated with the upper and lower inequality constraints respectively.
The Lagrangian is then given by
\begin{equation}
\begin{aligned}
    \mathcal{L}&(\rho, \nu, \lambda^+, \lambda^-) = \int_0^\infty d\omega \, \kappa(\omega)\rho(\omega) \\
    &\quad + \int_0^\infty d\omega \, \nu(\omega) \rho(\omega) \\
    &\quad - \sum_i \lambda_i^+ \left( \int_0^\infty d\omega \, \phi_i(\omega)\rho(\omega)
    - \overline\rho_i - \Delta\rho_i \right) \\
    &\quad + \sum_i \lambda_i^- \left( \int_0^\infty d\omega \, \phi_i(\omega)\rho(\omega)
    - \overline\rho_i + \Delta\rho_i \right) \,.
\end{aligned}
\end{equation}
Grouping the terms involving the spectral density $\rho(\omega)$, the Lagrangian can be rewritten as
\begin{equation}\label{eq:lagrangian_grouped}
\begin{aligned}
    &\mathcal{L}(\rho, \nu, \lambda^+, \lambda^-) =
    \sum_i (\lambda_i^+ - \lambda_i^-) \overline\rho_i + \sum_i (\lambda_i^+ + \lambda_i^-) \Delta\rho_i \\
    &+ \int_0^\infty d\omega \, \rho(\omega)
    \left[ \kappa(\omega) + \nu(\omega) - \sum_i (\lambda_i^+ - \lambda_i^-) \phi_i(\omega) \right] \,.
\end{aligned}
\end{equation}
Invoking the minimax theorem in the usual way, the optimal solution is obtained by noting that
\begin{equation}
    \sup_{\rho} \,\, \inf_{\nu(\omega), \lambda^\pm_i \geq 0} \mathcal{L}
    \leq \inf_{\nu(\omega), \lambda^\pm_i \geq 0} \,\, \sup_{\rho} \mathcal{L} \,,
\end{equation}
which allows the optimization to be performed first over the space of all permissible spectral densities (without any additional constraint, to be enforced by the Lagrange multipliers).

The supremum over the integral in \cref{eq:lagrangian_grouped} diverges unless the bracketed term evaluates to zero everywhere.
Therefore, a finite dual minimum only exists if
\begin{equation}
    \nu(\omega) = \kappa_\lambda(\omega) - \kappa(\omega) \,, \quad
    \kappa_\lambda(\omega) = \sum_i \lambda_i \phi_i(\omega) \,,
\end{equation}
where the unrestricted dual variable $\lambda_i \equiv \lambda_i^+ - \lambda_i^- \in \mathbb{R}$ was introduced.
For a fixed value of $\lambda_i$, the term proportional to $\Delta \rho_i \geq 0$ on the right-hand side \cref{eq:lagrangian_grouped} acts as a penalty. 
Since the dual formulation is cast as a minimization problem, for any fixed value of the difference $\lambda_i^+-\lambda_i^-\equiv \lambda_i$, the sum $\lambda_i^++\lambda_i^-$ must be minimized. Since both variables $\lambda_i^\pm$ are constrained to be non-negative, this minimum is achieved when at least one of them vanishes. 
Specifically, setting $\lambda_i^+ = \max(\lambda_i, 0)$ and $\lambda_i^- = \max(-\lambda_i, 0)$ immediately implies that $\lambda_i^+ + \lambda_i^- = |\lambda_i|$.
The minimization of the dual function then reduces exactly to the optimization problem presented in \cref{eq:dual-upper},
\begin{equation}
    \min_{\{\lambda_i \in \mathbb{R}\}} \,\, \sum_i \lambda_i \overline\rho_i + \sum_i |\lambda_i| \Delta\rho_i
\end{equation}
subject to the semi-infinite constraint
\begin{equation}
    \kappa_\lambda(\omega) = \sum_i \lambda_i \phi_i(\omega) \ge \kappa(\omega) \quad \forall \, \omega \in [0, \infty) \,.
\end{equation}
The $L_1$ absolute-value penalty emerges naturally as a direct analytic consequence of the $L_\infty$ box constraints on the input data.